\documentclass[%
 reprint,
 superscriptaddress,
 nofootinbib,
 amsmath,amssymb,
 aps,
 prd,
]{revtex4-2}

\usepackage[utf8]{inputenc}
\usepackage[english]{babel}
\usepackage{amsmath,latexsym}
\usepackage{csquotes}
\usepackage{graphicx}
\usepackage{xcolor}
\usepackage[colorlinks=true,linkcolor=magenta,citecolor=blue]{hyperref}
\usepackage{subcaption}   

\DeclareCaptionJustification{fulljust}{\leftskip=0pt\rightskip=0pt\parfillskip=0pt plus 1fil\relax}
\begin{document}

\title{Rotating Neutron Star Migrations as a Standardized Test for 3+1 Numerical Relativity}

\author{Óscar H. Petit}
\affiliation{Departamento de Astronom\'{\i}a y Astrof\'{\i}sica, Universitat de Val\`encia, Avinguda Vicent 
Andrés Estellés 19, 46100, Burjassot (Val\`encia), Spain.}
\email{oscar.hervas@uv.es}

\author{José A. Font}
\affiliation{Departamento de Astronom\'{\i}a y Astrof\'{\i}sica, Universitat de Val\`encia, Avinguda Vicent 
Andrés Estellés 19, 46100, Burjassot (Val\`encia), Spain.}
\affiliation{Observatori Astron\`omic, Universitat de Val\`encia, C/ Catedr\'atico Jos\'e Beltr\'an 2, 46980, Paterna (Val\`encia), Spain.}

\author{Nikolaos Stergioulas} 
\affiliation{Department of Physics, Aristotle University of Thessaloniki, Thessaloniki 54124, Greece}

\date{September 18, 2026}

\begin{abstract}
When simulating dynamically unstable neutron stars, truncation errors
can introduce perturbations that drive the stellar configuration either
toward gravitational collapse into a black hole or toward migration to
a dynamically stable, lower-density configuration. This mechanism has
become a standard benchmark for validating general-relativistic
hydrodynamics codes in the case of spherically symmetric neutron-star
models. In this work, we extend the analysis to the more general case
of uniformly rotating neutron stars evolved in three spatial dimensions. We
investigate these transitions using two equations of state: a cold
polytrope and a hybrid prescription, which allow us to distinguish the
barotropic response from the effects of shock-generated thermal
pressure. We find that the migrated remnant remains close to uniform
rotation after the transition, and we trace its quasi-radial pulsations
through complementary diagnostics involving the central density, the
rotation profile, and the axisymmetric gravitational-wave channel. We
investigate the properties of these transitions and propose a
standardized test for validating numerical-relativity codes by
requiring the migration behavior to be consistent with the dynamics
reported here.
\end{abstract}

\maketitle

\section{Introduction}
Neutron stars (NSs) are among the most extreme astrophysical laboratories accessible to direct observation, hosting matter at densities exceeding nuclear saturation density and probing regions of the quantum chromodynamics phase diagram that remain inaccessible to experiment~\cite{Lattimer:2012_review, Oertel:2017_RMP}.
The detection of gravitational waves from binary neutron-star
mergers~\cite{LIGO:2017_GW170817}, together with their
electromagnetic counterparts~\cite{LIGO:2017_multimessenger},
has firmly established multi-messenger astronomy as a tool for
constraining the equation of state (EoS) of dense matter, the engines
of short gamma-ray bursts, and the sites of heavy-element
nucleosynthesis~\cite{Metzger:2019_kilonova_review}.

The theoretical interpretation of these observations relies heavily on numerical-relativity simulations of the coupled Einstein--Euler system. Such simulations must accurately capture phenomena spanning many orders of magnitude in length scale and density: strong shocks, ultrarelativistic outflows, magnetized turbulence, and, increasingly, neutrino transport, all while preserving the constraints of general relativity over many dynamical timescales. Validating the sophisticated numerical-relativity codes used for these simulations
is an essential ingredient of the scientific pipeline. Over the past two decades, a canonical suite of benchmark problems has been established to address this need, including one-dimensional shock-tube tests, calculations of the frequency spectra of radial and non-radial oscillation modes of relativistic stars, long-term evolutions of stable Tolman--Oppenheimer--Volkoff (TOV) configurations, and convergence studies and code comparisons for assessing the numerical accuracy of binary neutron star merger simulations and their predicted gravitational waveforms (see e.g.~\cite{Marti:2003_review, Font:2008_review, Baiotti:2010,Bernuzzi:2012,Read:2013,Radice:2014,Baiotti:2017_NSreview,Paschalidis:2017,Kiuchi:2017,Babiuc:2025,Habib:2026, Siebel_2002} and references therein).

Among these benchmarks, the \emph{migration} of an unstable neutron star toward a dynamically stable equilibrium configuration has acquired a particularly important
role~\cite{Font:2001_grqc_0110047,Font:2002_3Dtests,Cordero-Carrion:2009,Bernuzzi:2010,Thierfelder:2011,Galeazzi_2013,Cheong:2021,Camelio:2023,Lam:2025}. The test stresses essentially every component of a general-relativistic hydrodynamics code: the conservative-to-primitive recovery scheme must remain robust under
order-of-magnitude variations of the central density; the high-resolution shock-capturing (HRSC) algorithm must resolve the strong shocks generated by the rebounding stellar surface; and the spacetime evolution must remain stable while the gravitational well oscillates by significant fractions of its
initial depth. Importantly, the qualitative outcome, migration
versus prompt collapse to a black hole, is sharply sensitive to the
truncation error of the underlying scheme, providing a strong and
visually unambiguous diagnostic of code
fidelity~\cite{Cordero-Carrion:2009}.

The dynamical stability of a non-rotating neutron star against radial
perturbations is determined by the fundamental radial mode. Along a
one-parameter sequence of TOV equilibria, its squared eigenfrequency
decreases as the maximum-mass configuration is approached and vanishes
at the neutral-stability point,
$\omega_F^2=0.$
For a barotropic non-rotating star, this point coincides with the
maximum-mass turning point of the equilibrium sequence. Configurations
on its higher-density side are dynamically unstable: depending on the
applied perturbation, they either collapse to a black hole or migrate
toward a stable configuration with the same baryonic
mass~\cite{Sorkin:1981_turning,Sorkin:1982_turning,
Font:2001_grqc_0110047}. In numerical evolutions, discretization errors
can supply the perturbation that triggers either outcome.

Rotation separates the dynamical neutral-stability point from the
turning point. Uniformly rotating stars form a two-parameter family of
equilibria that may be labeled by the central density $\rho_c$ and
angular momentum $J$, or equivalently by another rotational parameter
such as $\Omega$ or $r_p/r_e$~\cite{Paschalidis:2017}. Along a
constant-$J$ sequence, the turning point is conventionally defined by
\begin{equation}
\label{eq:rot_turningpoint}
\left.\frac{\partial M}{\partial\rho_c}\right|_J=0,
\end{equation}
where $M$ is the gravitational mass. The turning-point theorem provides
a sufficient condition for secular instability, but it does not locate
the exact onset of dynamical instability in a rotating
star~\cite{Friedman:1988_turningpoint,Friedman:2013_book}. The latter is
instead determined by the neutral-stability line on which the
quasi-radial fundamental mode satisfies $\omega_F^2=0$.

Full general relativistic calculations show that, for uniformly rotating $K=100, \ \Gamma=2$ polytropic stars along sequences of increasing
central density, the rotating neutral-stability line is encountered
before the turning-point line, with their separation increasing with
rotation~\cite{Takami:2011_rotNS_stability}. Consequently, some
configurations on the lower-density side of the turning point are
already dynamically unstable. This implies that every configuration located
on the higher-density side of the turning point also lies beyond the
neutral-stability line. All initial models evolved in this work are selected on the higher-density
side of their corresponding constant-$J$ turning points
(Sec.~\ref{subsec:id_grid}), and are separated from those turning points by a
margin in central density that is large compared with the offset between the
turning-point and neutral-stability lines reported in
Ref.~\cite{Takami:2011_rotNS_stability}. The models are thus well inside the
dynamically unstable region, and the migration dynamics we discuss is
insensitive to the precise location of the neutral-stability line. Since no
closed-form criterion exists for the latter, whereas the turning point follows
directly from the equilibrium sequences themselves, we adopt the turning point
throughout as a convenient and conservative reference, keeping in mind that
the physically relevant boundary is the dynamical one.

Rotating neutron-star migration has received considerably less attention than its non-rotating counterpart. Previous axisymmetric studies investigated migrations triggered by a phase transition induced instability and included rotating migration and collapse tests ~\cite{Dimmelmeier:2009_rotating_migration,Lam:2025}. In this work, we extend these investigations to a controlled sequence of uniformly rotating models evolved in full three dimensions and formulate their migration dynamics as a standardized benchmark for numerical-relativity codes.

Rotation is also astrophysically relevant. Pulsar timing has revealed
spin frequencies up to $716 \, \mathrm{Hz}$~\cite{Hessels:2006_J1748},
while centrifugal support in binary-neutron-star merger remnants can
temporarily sustain masses above the non-rotating maximum before delayed
collapse~\cite{Baumgarte:2000_supramassive,
Hotokezaka:2013_remnants}.

Whether a full 3D 3+1 numerical-relativity code's response to perturbations on this rotating unstable branch is well-behaved is, to our knowledge, untested as a community benchmark. The dynamics in the rotating case are qualitatively richer than in the spherical one. In addition to the radial pulsations, the system can
redistribute angular momentum via
shock-heated envelope expansion and develop non-axisymmetric (bar-mode) deformations on
either dynamical or secular timescales~\cite{Paschalidis:2017,
Loeffler:2014_bar_review,Manca:2007_bar}. These additional degrees of freedom render
the test substantially more demanding and, simultaneously, more
representative of the post-merger regime in which contemporary
numerical-relativity codes are most frequently deployed.

In this work, we study the migration of uniformly rotating neutron
stars through full three-dimensional evolutions of the coupled
Einstein--Euler system. The hydrodynamic sector is evolved using the
\texttt{GRHayL} infrastructure~\cite{Cupp:2025grhayl}, while the
equilibrium initial data are generated with the \texttt{RNS}
code~\cite{Stergioulas:1995_RNS}. We consider dynamically unstable
models spanning different central densities and rotation rates and
follow their migration toward lower-density configurations. We
characterize: (i) their trajectories on the normalized
baryonic mass--central energy density plane,
$\bar{M}_0$--$\bar{\epsilon}_c$, in relation to the corresponding
constant-angular-momentum equilibrium sequences; (ii) the time-domain
migration dynamics for cold and hybrid equations of state, including
the central-density evolution, rebound-shock structure, shock-mediated
damping and evolution of the rotation profile; and (iii) the
gravitational-wave emission accompanying the transition. Based on these results, we
propose a standardized qualitative benchmark for assessing
three-dimensional numerical-relativity codes in the rotating regime.

The remainder of the paper is organized as follows.
Section~\ref{sec:basic_equations} presents the $3+1$ formalism, the
\texttt{GRHayL} conservative formulation of relativistic
hydrodynamics, and the numerical methods employed in our evolutions.
Section~\ref{sec:results} presents our results: the location of the
evolved models on the $\bar{M}_0$--$\bar{\epsilon}_c$ plane, the
time-domain dynamics of three hybrid models, and the gravitational-wave
signal. Section~\ref{sec:conclusions} summarizes our findings and lays out the proposed standardized test. Throughout this paper we adopt
geometrized units in which $G = c = M_\odot = 1$, with conversions to physical units indicated where appropriate. Greek indices run over spacetime components $\{0,1,2,3\}$ and Latin indices over spatial components $\{1,2,3\}$; we use the metric signature $(-,+,+,+)$ and Einstein's summation convention.

\section{Mathematical Framework and Numerical Methodology}
\label{sec:basic_equations}

To investigate the dynamical migration of unstable rotating neutron stars to a stable equilibrium state, we must solve the fully coupled Einstein-Euler system. The migration process is characterized by violent radial pulsations, expansion/contraction, and severe shock heating \cite{Font:2001_grqc_0110047}. Resolving these highly nonlinear dynamics without succumbing to numerical instabilities requires robust computational infrastructure. For the hydrodynamics sector, we rely on the General Relativistic Hydrodynamics Library (\texttt{GRHayL}) \cite{Cupp:2025grhayl}, which provides  infrastructure-agnostic flux calculations, high-resolution shock-capturing (HRSC) schemes, and highly resilient conservative-to-primitive recovery methods necessary to track the star through its extreme transitional phases.

\subsection{Spacetime Evolution: The 3+1 Formalism}
\label{subsec:3plus1}

The evolution of the gravitational field is treated within the standard $3+1$ 
decomposition of general relativity. The four-dimensional spacetime manifold $\mathcal{M}$ with metric $g_{\mu\nu}$ is foliated into a family of spacelike hypersurfaces $\Sigma_t$, parameterized by a global time coordinate $t$. In addition, $n^\mu$ denotes the future-pointing, timelike unit normal vector to the hypersurfaces $\Sigma_t$.

The spacetime line element is expressed as:
\begin{equation}
    ds^2 = g_{\mu\nu} dx^\mu dx^\nu = -\alpha^2 dt^2 + \gamma_{ij} (dx^i + \beta^i dt)(dx^j + \beta^j dt),
\end{equation}
where $\alpha$ is the lapse function, $\beta^i$ is the spacelike shift vector, and $\gamma_{ij}$ is the spatial metric induced on $\Sigma_t$. The embedding of $\Sigma_t$ within $\mathcal{M}$ is described by the extrinsic curvature tensor $K_{ij}$. The evolution of the spacetime is governed by the Einstein field equations, which decompose into the Hamiltonian and momentum constraint equations, and the evolution equations for $\gamma_{ij}$ and $K_{ij}$. The \texttt{GRHayL} framework integrates cleanly with the metric evolution by taking $\alpha, \beta^i, \gamma_{ij}$, and their spatial derivatives as inputs to construct the geometric source terms that govern the fluid dynamics \cite{Cupp:2025grhayl}.

\subsection{General Relativistic Hydrodynamics: The GRHayL Formulation}
\label{subsec:grhd}

The neutron star matter is approximated as an ideal perfect fluid. Its dynamics are governed by the local conservation of baryon number, $\nabla_\mu (\rho u^\mu) = 0$, and the conservation of energy-momentum, $\nabla_\mu T^{\mu\nu} = 0$. The stress-energy tensor for a perfect fluid is:
\begin{equation}
    T^{\mu\nu} = \rho h u^\mu u^\nu + P g^{\mu\nu},
\end{equation}
where $\rho$ is the rest-mass density, $P$ is the fluid pressure, $u^\mu$ is the fluid four-velocity, and $h = (\epsilon+P)/\rho=1 + e + P/\rho$ is the specific relativistic enthalpy, with $e$ representing the specific internal energy, and the total energy density $\epsilon = \rho(1+e)$.

To reliably evolve the fluid in the presence of strong shocks, the equations of motion are cast into a strongly hyperbolic, flux-conservative system. While many numerical-relativity codes use the standard Valencia formulation (which defines the fluid velocity relative to an Eulerian observer)~\cite{Valencia}, \texttt{GRHayL} inherits a highly robust variant from the \texttt{IllinoisGRMHD} code \cite{Cupp:2025grhayl}. In the \texttt{GRHayL} formulation, the primitive fluid variables are
written as $\mathbf{V}=(\rho,P,\tilde v^i)$, where
\begin{equation}
    \tilde v^i \equiv \frac{u^i}{u^t}
                    = \frac{dx^i}{dt}
\end{equation}
is the coordinate, or transport, velocity. We use a tilde to distinguish
this quantity from the standard Eulerian three-velocity $V^i$ measured
by observers normal to the spatial hypersurfaces. The two velocities
are related by
\begin{equation}
    \tilde v^i = \alpha V^i-\beta^i,
    \qquad
    V^i=\frac{\tilde v^i+\beta^i}{\alpha}.
\end{equation}
The corresponding Lorentz factor is
\begin{equation}
    W=\alpha u^t=
    \left(1-\gamma_{ij}V^iV^j\right)^{-1/2}.
\end{equation}
Equivalently, $u^t$ can be expressed directly in terms of the transport
velocity as
\begin{equation}
    u^t =
    \left[
      \alpha^2-
      \gamma_{ij}(\tilde v^i+\beta^i)
                    (\tilde v^j+\beta^j)
    \right]^{-1/2}.
\end{equation}

The corresponding state vector of conservative variables $\mathbf{U} = (\rho_*, \tilde{S}_j, \tilde{\tau})$ represents the conserved rest-mass density, momentum density, and energy density. In \texttt{GRHayL}, these are defined in terms of the stress-energy tensor components as
\begin{eqnarray}
    \rho_* &=& \alpha \sqrt{\gamma} \rho u^0, \label{eq:cons_rhostar} \\
    \tilde{S}_j &=& \alpha \sqrt{\gamma} T^0_{\ j}, \label{eq:cons_S} \\
    \tilde{\tau} &=& \alpha^2 \sqrt{\gamma} T^{00} - \rho_*. \label{eq:cons_tau}
\end{eqnarray}
where $\gamma = \det(\gamma_{ij})$. The fluid evolution equations are then written in the characteristic balance-law form:
\begin{equation}
    \partial_t \mathbf{U} + \partial_i \mathbf{F}^i = \mathbf{S}.
\end{equation}
The flux vectors $\mathbf{F}^i$, taking advantage of the coordinate velocity definition, are:
\begin{equation}
\mathbf{F}^i =
    \begin{pmatrix}
    \rho_*\tilde v^i \\
    \alpha\sqrt{\gamma}T^{i}{}_{j} \\
    \alpha^2\sqrt{\gamma}T^{0i}-\rho_*\tilde v^i
    \end{pmatrix}.
\end{equation}
The source term vector $\boldsymbol{S}$ encapsulates the geometric coupling between the fluid and the curved spacetime,
\begin{equation}
    \boldsymbol{S} = 
    \begin{bmatrix}
        0 \\
        \frac{1}{2}\alpha\sqrt{\gamma}T^{\mu\nu}g_{\mu\nu,i}\\
        \alpha\sqrt{\gamma} \left[ \Theta^{kl} K_{kl} - \left(T^{00}\beta^k + T^{0k}\right)\partial_k \alpha \right]
    \end{bmatrix},
\end{equation}
where $\Theta^{kl} \equiv T^{00}\beta^k\beta^l + 2T^{0k}\beta^l + T^{kl}$. The \texttt{GRHayL} library isolates the calculation of these fluxes and geometric sources into modular routines, ensuring precise discrete conservation properties even during the violent structural transitions of the migrating star \cite{Cupp:2025grhayl}.

\subsection{Spacetime Evolution Equations: BSSN Formulation}
\label{subsec:bssn}

The 3+1 Arnowitt-Deser-Misner (ADM) equations in their original form are only weakly hyperbolic and unsuitable for stable, long-term numerical evolution. We therefore adopt the BSSN reformulation~\cite{Shibata:1995_BSSN,Baumgarte:1998_BSSN}, which recasts the field equations into a strongly
hyperbolic system via a conformal decomposition of the spatial metric
and a trace/traceless split of the extrinsic curvature. The conformal
factor $\phi$ and conformal spatial metric $\tilde{\gamma}_{ij}$ are
defined by
\begin{equation}
    \tilde{\gamma}_{ij} = e^{-4\phi}\gamma_{ij},
    \qquad
    \det\tilde{\gamma}_{ij} = 1,
\end{equation}
while the extrinsic curvature is decomposed as
\begin{equation}
    K_{ij} = e^{4\phi}\!\left(\tilde{A}_{ij}
            + \tfrac{1}{3}\tilde{\gamma}_{ij}K\right),
    \qquad
    \tilde{\gamma}^{ij}\tilde{A}_{ij} = 0,
\end{equation}
where $K = \gamma^{ij}K_{ij}$. The conformal connection functions
\begin{equation}
    \tilde{\Gamma}^i \equiv -\partial_j \tilde{\gamma}^{ij}
\end{equation}
are promoted to independent evolved variables. The resulting evolution
system for
$\{\phi,\, \tilde{\gamma}_{ij},\, K,\, \tilde{A}_{ij},\, \tilde{\Gamma}^i\}$
displays the long-term stability required to follow unstable
configurations through their violent rebound phase.

The coordinate freedom of the $3+1$ formulation is fixed through the
``moving-puncture'' gauge~\cite{Campanelli:2006_MP, Baker:2006_MP},
which combines the $1+\log$ slicing
condition~\cite{Bona:1995_slicing} for the lapse with a hyperbolic
$\Gamma$-driver shift~\cite{Alcubierre:2003_gammadriver}. The lapse
evolves according to
\begin{equation}
    \partial_t \alpha = \beta^i \partial_i \alpha - 2\,\alpha\,K,
    \label{eq:gauge_alpha}
\end{equation}
and the shift vector is advanced through the auxiliary field $B^i$ as
\begin{align}
    \partial_t \beta^i &= \beta^j \partial_j \beta^i + \tfrac{3}{4}\,B^i,
    \label{eq:gauge_beta} \\
    \partial_t B^i     &= \beta^j \partial_j B^i
                       + \partial_t \tilde{\Gamma}^i
                       - \beta^j \partial_j \tilde{\Gamma}^i
                       - \eta\, B^i,
    \label{eq:gauge_B}
\end{align}
with damping coefficient $\eta = 1$ (in code units). In practice the
equations are integrated with the \texttt{ML\_BSSN}
thorn~\cite{Brown:2009_McLachlan} of the \texttt{Einstein
Toolkit}~\cite{Loeffler:2012_ET}, with a minimum-lapse floor
$\alpha_{\min} = 10^{-8}$ to prevent unphysical zero-crossings during the bounce. This gauge choice is singularity-avoiding and, together with the BSSN evolution system, allows the central lapse to plunge and rebound during the migration without driving coordinate pathologies.

\subsection{Equation of State}
\label{subsec:eos}

The migration dynamics are sensitive to the thermodynamic prescription
adopted for the stellar matter. To disentangle purely hydrodynamical
effects from those driven by shock heating, we employ two complementary
equations of state.

\paragraph{Cold polytropic EoS.} As a barotropic baseline we use
\begin{equation}
    P = K \rho^{\Gamma},
    \qquad
    e = \frac{K \rho^{\Gamma-1}}{\Gamma - 1},
    \label{eq:eos_cold}
\end{equation}
with polytropic constant $K=100$ and adiabatic index
$\Gamma = 1 + 1/n = 2$ ($n=1$). The same EoS is enforced during the
evolution so that no entropy is generated and the migration proceeds
adiabatically.

\paragraph{Hybrid (cold $+$ thermal) EoS.} To incorporate shock heating
we adopt the standard hybrid prescription~\cite{Janka:1993_hybridEoS}
\begin{equation}
    P = P_{\rm cold}(\rho) + P_{\rm th},
    \qquad
    P_{\rm th} = (\Gamma_{\rm th} - 1)\,\rho\,e_{\mathrm{th}},
    \label{eq:eos_hybrid}
\end{equation}
where $P_{\rm cold}$ follows Eq.~\eqref{eq:eos_cold} and the thermal
specific internal energy is
$e_{\mathrm{th}} = e-e_{\mathrm{cold}}(\rho)$. We adopt
$\Gamma_{\rm th} = 2$. Initial data are built with the cold EoS in both
cases; the hybrid form acts only during the evolution.

\subsection{Numerical Methods}
\label{subsec:numerics}

High-resolution shock-capturing of the fluid equations is provided by
\texttt{GRHayL}~\cite{Cupp:2025grhayl}. Primitive variables are
reconstructed at cell interfaces with the piecewise parabolic method
(PPM)~\cite{Colella:1984_PPM}, and intercell numerical fluxes are
computed with the HLLE approximate Riemann
solver~\cite{Harten:1983_HLLE, Einfeldt:1988_HLL}. The semi-discrete
system is advanced in time by the method of lines using a fourth-order
Runge--Kutta integrator with Courant factor $0.15$. The \texttt{GRHayL} infrastructure has previously been validated using
shock-tube tests, stable TOV evolutions, and binary neutron-star
simulations~\cite{Cupp:2025grhayl}. In particular, stable TOV
evolutions showed an approximate convergence order of
$n\simeq2.5$ for the central-density error.

The conservative-to-primitive (con2prim) inversion uses the
one-dimensional scheme of Palenzuela
et~al.~\cite{Palenzuela:2015_con2prim} as implemented in
\texttt{GRHayL}. In evolutions employing the cold EoS, we use the barotropic relation $e=e_{\mathrm{cold}}(\rho)$
where no thermal component is present.

Outside the stellar surface the rest-mass density is set to a tenuous
artificial atmosphere with a floor value
$\rho_{\rm atm} = 10^{-11}$ in geometrized units. Cells whose recovered
density falls below this threshold are reset to atmosphere values, with
the fluid three-velocity set to zero in the Eulerian frame. The
atmosphere level is chosen well below the densities reached by the
ejected envelope during the migration bounce, to minimize its influence on the bulk evolution.

All migrations are triggered by finite-resolution truncation errors
rather than by an explicitly imposed perturbation. Since changing the
resolution also changes this perturbation, the benchmark concerns the
migration dynamics rather than its precise onset time.

\begin{table*}[t]
\centering
\caption{Initial parameters of the rotating neutron-star models evolved in this work. Here, \(\rho_c\) is the central rest-mass density, \(r_p/r_e\) the polar-to-equatorial coordinate-axis ratio, \(\bar{J}\) the normalized angular momentum, \(M\) the gravitational mass, \(M_0\) the baryonic mass, \(T/W\) the ratio of rotational kinetic to gravitational binding energy, \(\Omega\) the angular velocity, and \(R_e\) the equatorial radius. The Hybrid-EoS family employs the hybrid prescription of Eq.~\eqref{eq:eos_hybrid}, the constant-\(\rho_c\) family employs the cold barotropic EoS of Eq.~\eqref{eq:eos_cold} and the constant-\(\bar J\) family is evolved with both EoS.}
\label{tab:initial_data}
\setlength{\tabcolsep}{4pt}
\small
\begin{tabular}{l l c c c c c c c c}
\hline\hline
Family & Model & $\rho_c$ & $r_p/r_e$ & $\bar{J}$ & $M$ & $M_0$
 & $T/W$ & $\Omega$ & $R_e$ \\
 & & $[\times 10^{-3}]$ & & & $[M_\odot]$ & $[M_\odot]$
 & $[\times 10^{-2}]$ & $[\times 10^{-2}M_\odot^{-1}]$ & $[M_\odot]$\\
\hline
Hybrid-EoS
  & H1 & 8.00 & 0.95 & 0.003 & 1.458 & 1.546 & 0.586 & 1.924 & 5.907 \\
  & H2 & 8.00 & 0.80 & 0.009 & 1.527 & 1.617 & 3.842 & 4.710 & 6.452 \\
  & H3 & 6.00 & 0.70 & 0.014 & 1.684 & 1.813 & 5.961 & 4.963 & 7.541 \\
\hline
\shortstack[l]{Cold-EoS (const.\ $\rho_c$)}
  & C1 & 4.70 & 1.00 & 0     & 1.597 & 1.744 & 0     & 0     & 6.812 \\
  & C2 & 4.70 & 0.94 & 0.005 & 1.628 & 1.776 & 1.339 & 2.222 & 6.987 \\
  & C3 & 4.70 & 0.84 & 0.010 & 1.676 & 1.829 & 3.364 & 3.437 & 7.311 \\
  & C4 & 4.70 & 0.71 & 0.015 & 1.755 & 1.914 & 6.205 & 4.480 & 8.050 \\
\hline
\shortstack[l]{Cold/Hybrid-EoS (const.\ $\bar{J}$)}
  & J1 & 3.88 & 0.85 & 0.010 & 1.712 & 1.877 & 3.492 & 3.177 & 7.748 \\
  & J2 & 5.66 & 0.83 & 0.010 & 1.636 & 1.771 & 3.652 & 3.918 & 6.987 \\
  & J3 & 6.84 & 0.80 & 0.010 & 1.586 & 1.698 & 4.048 & 4.490 & 6.735 \\
\hline\hline
\end{tabular}
\end{table*}

\subsection{Initial Data and Computational Setup}
\label{subsec:id_grid}

Equilibrium rotating neutron-star initial data are generated with the
\texttt{RNS} code~\cite{Stergioulas:1995_RNS}, which solves the
Einstein equations for stationary, axisymmetric, uniformly rotating
perfect-fluid configurations on a spherical grid and is subsequently
interpolated onto our Cartesian numerical domain. All initial data are
built with the cold polytropic EoS of Eq.~\eqref{eq:eos_cold}.

To probe the migration dynamics we evolve
three families of models.

\paragraph{Hybrid-EoS.} Three configurations are evolved with the
hybrid EoS of Eq.~\eqref{eq:eos_hybrid}: two with central rest-mass
density $\rho_c = 8\times 10^{-3}$ and polar-to-equatorial axis ratios
$r_p/r_e = 0.95$ and $0.80$, and a third one with $\rho_c = 6\times 10^{-3}$
and $r_p/r_e = 0.70$.

\paragraph{Cold-EoS.} For the purely barotropic evolutions we
consider two complementary sequences. A constant--$\rho_c$ sequence
fixes $\rho_c = 4.7\times 10^{-3}$ and samples axis ratios
$r_p/r_e \in \{1.00,\, 0.94,\, 0.84,\, 0.71\}$, isolating the effect of
rotation at fixed central compactness. A constant--$\bar{J}$ sequence
fixes the $K$-scaled angular momentum
$\bar{J} \equiv J/K^n = 0.01$ (with $n=1$) and samples central densities
$\rho_c \in \{3.88,\, 5.66,\, 6.84\}\times 10^{-3}$, scanning the
unstable branch at fixed total angular momentum. The three constant-$\bar{J}$ models J1--J3 were additionally re-evolved
with the hybrid EoS of Eq.~\eqref{eq:eos_hybrid} from identical initial
data, so that the thermal contribution to the migration endpoint can be
isolated at fixed baryonic mass and angular momentum
(Sec.~\ref{subsubsec:thermal_offset}).
The full set of
initial parameters is summarized in Table \ref{tab:initial_data}.

We introduce the $K$-scaled baryonic mass $\bar{M_0}\equiv M_0/\sqrt{K}$, and energy density $\bar{\epsilon}=K\epsilon$ together with the already introduced $\bar J$ for a better visualization in the following sections.

All evolutions are performed on a Cartesian domain of half-width
$144$ in code units, using five nested refinement levels, including
the coarsest level, managed by \texttt{Carpet}
\cite{Schnetter:2004_Carpet}. The coarsest level has uniform grid
spacing $\Delta x=\Delta y=\Delta z=6.0$, and each successive level
increases the resolution by a factor of two. The grid spacings are
therefore
\begin{equation}
    \Delta x_\ell =
    (6.0,\,3.0,\,1.5,\,0.75,\,0.375),
    \qquad \ell=0,\ldots,4,
\end{equation}
giving a finest resolution of $\Delta x_{\rm fin}=0.375$. The four
refined boxes have outer radii
\begin{equation}
    R_\ell =
    (84.0,\,66.0,\,42.0,\,30.0),
    \qquad \ell=1,\ldots,4.
\end{equation}

\section{Results}
\label{sec:results}
We organize the discussion in four parts. We first locate every evolved cold model on the $\bar{M}_0$--$\bar{\epsilon}_c$ plane and examine whether
the cold-EoS migration proceeds along the constant-$\bar{J}$
equilibrium sequence associated with the initial angular momentum. We then analyze the evolution of the central
density, contrasting the two EoS: the thermal offset of
the migration endpoint along the constant-$\bar{J}$ sequence and the
frequency content of the post-bounce quasi-radial pulsations. We next
dissect the time-domain dynamics of the three hybrid configurations, H1
($\rho_c=8\times 10^{-3}$, $r_p/r_e=0.95$), H2 ($\rho_c=8\times
10^{-3}$, $r_p/r_e=0.80$) and H3 ($\rho_c=6\times 10^{-3}$,
$r_p/r_e=0.70$), highlighting the qualitative differences in their shock
structure and angular-momentum redistribution. Finally we present the
gravitational-wave analysis of the transition.

\subsection{Migration in the $\bar{M}_0$--$\bar{\epsilon}_c$ plane}
\label{subsec:Meps}

\begin{figure}[tbp]
    \centering
    \includegraphics[width=\linewidth]{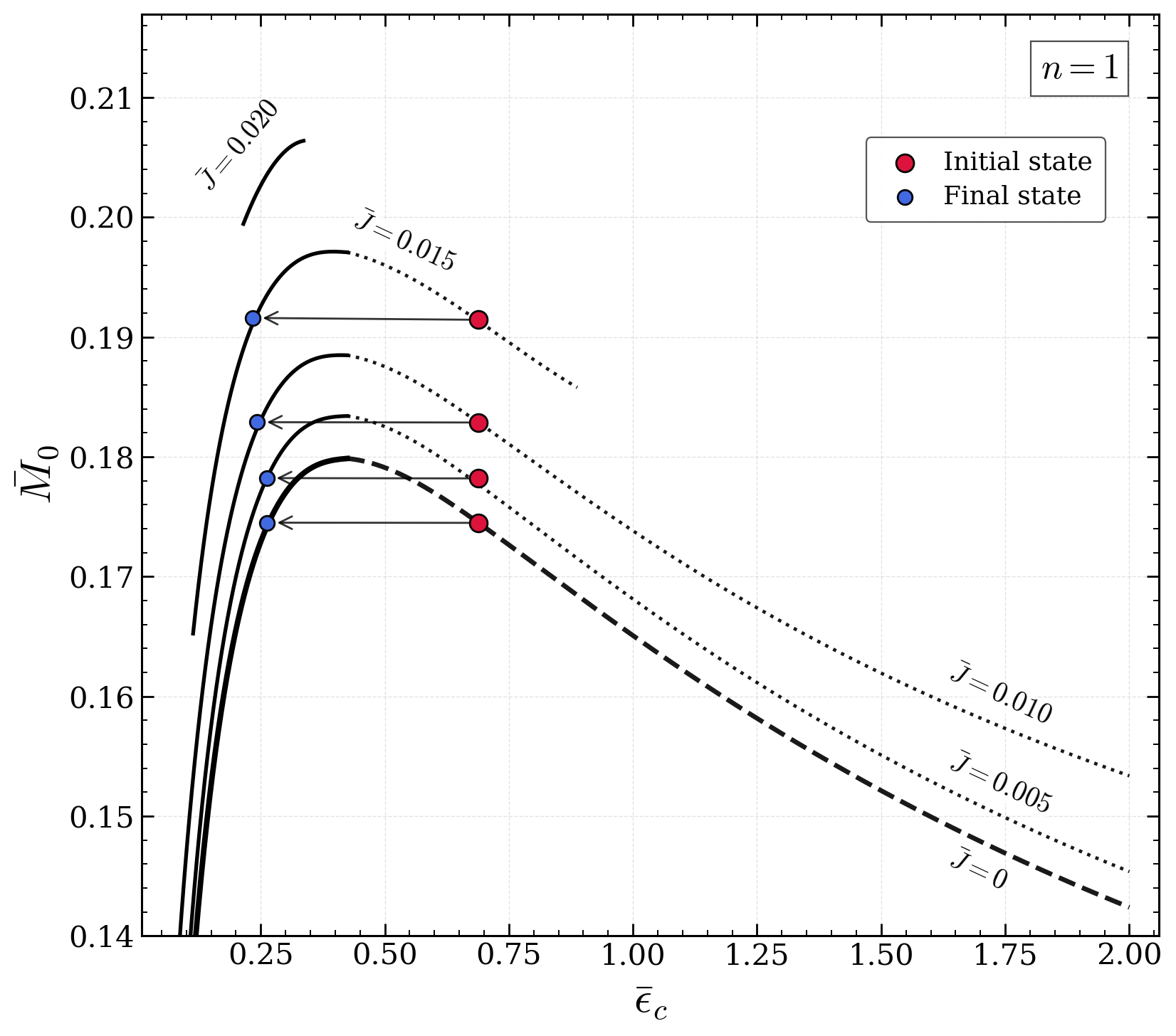}
    \caption{Migration trajectories on the
    $\bar{M}_0$--$\bar{\epsilon}_c$ plane for the cold-EoS
    constant-$\rho_c$ family ($\rho_c=4.7\times10^{-3}$). Solid and dotted
    curves denote, respectively, the lower- and higher-density portions of
    constant-$\bar{J}$ equilibrium sequences separated by their turning
    points; the dashed curve denotes the non-rotating limit.}
    \label{fig:Meps_constrho}
\end{figure}

\begin{figure}[tbp]
    \centering
    \includegraphics[width=\linewidth]{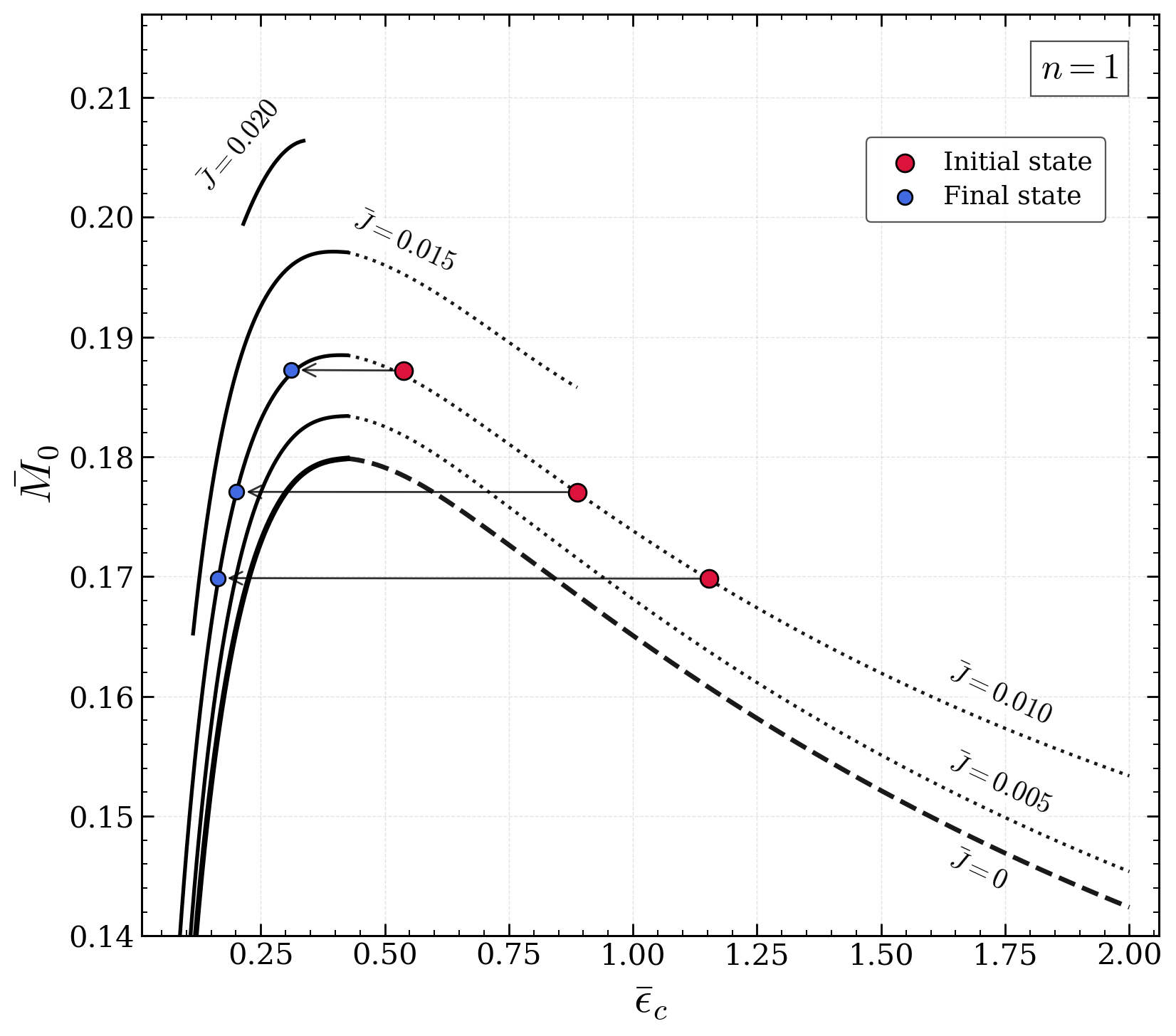}
    \caption{Same as Fig.~\ref{fig:Meps_constrho}, but for models with
    initial angular momentum $\bar{J}=0.01$.}
    \label{fig:Meps_constJ}
\end{figure}

Figures~\ref{fig:Meps_constrho} and \ref{fig:Meps_constJ} summarize the
global outcome of the cold-EoS migrations on the $K$-scaled mass--energy
diagram. The solid and dotted portions of each
constant-$\bar{J}$ equilibrium sequence denote, respectively, its
lower and higher density sides relative to the turning point. Red dots mark the initial
(dynamically unstable) configurations; blue dots mark the final states
reached after the system has relaxed.

The constant-$\rho_c$ family in Fig.~\ref{fig:Meps_constrho} fixes
$\rho_c = 4.7\times 10^{-3}$ and increases the rotation rate along the
sequence. Each model migrates towards lower $\bar{\epsilon}_c$ and settles
on the less-dense state of the constant-$\bar{J}$ curve corresponding to
its initial angular momentum, as required by the conservation
of $J$ in the cold evolutions. As rotation increases, the migrated configurations occur at
progressively larger $\bar{M}_0$, reflecting the additional
centrifugal support.

The constant-$\bar{J}$ family in Fig.~\ref{fig:Meps_constJ} fixes
$\bar{J}=0.01$ and instead varies the initial central density. The three models migrate to the same
$\bar{J}=0.01$ curve, with the most compact initial configuration
producing the largest path in $\bar{\epsilon}_c$. This behavior provides a
diagnostic of angular-momentum conservation.

\subsection{Central density evolution}
\label{rhocsection}

\begin{figure}[tbp]
    \centering
    \includegraphics[width=\linewidth]{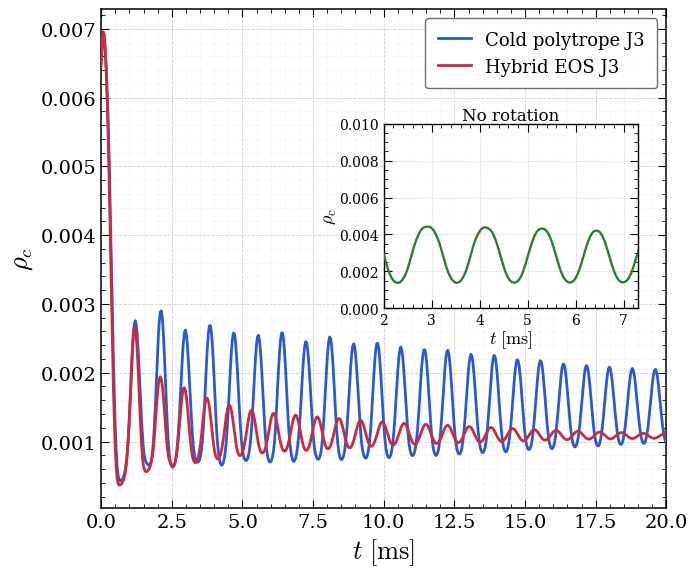}
    \caption{Central rest-mass density $\rho_c$ as a function of time for the J3
    configuration evolved with the cold $\Gamma=2$ polytrope (blue) and the
    hybrid cold$+$thermal EoS (red), starting from the same unstable initial data
    ($\rho_c\simeq 6.8\times 10^{-3}$). \emph{Inset:}
    Non-rotating, less dense cold model C1.}
    \label{fig:rhoc}
\end{figure}

A diagnostic of the migration is the time evolution of the central rest-mass
density $\rho_c(t)$, which compresses the global dynamics of the transition into
a single scalar evolution. Figure~\ref{fig:rhoc} shows $\rho_c(t)$ for the J3
configuration of Table~\ref{tab:initial_data} ($\rho_c(0)\simeq 6.8\times
10^{-3}$), evolved with each of our two equations of
state: the cold $\Gamma=2$ polytrope (blue) and the hybrid cold$+$thermal
prescription (red). Both runs share identical initial data, so any difference in
the subsequent evolution isolates the effect of shock heating.

Both models display the same initial behavior. Starting from the unstable equilibrium configuration, the star expands rapidly and its central density decreases by more than an order of magnitude, reaching $\rho_c\simeq4\times10^{-4}$ at $t\simeq0.5 \ \mathrm{ms}$. The expansion subsequently reverses, initiating a sequence of large-amplitude quasi-radial pulsations. This initial expansion-and-rebound behavior is common to both prescriptions because they coincide before appreciable shock heating develops.

The two evolutions diverge sharply thereafter. In the cold run the barotropic
EoS ties the pressure rigidly to the density, so the kinetic energy of each
bounce cannot be converted into heat; lacking this dissipation channel, the
radial pulsations remain large and only weakly damped, persisting with nearly
constant period throughout the full $20\,$ms shown and still reaching $\rho_c\sim
2\times 10^{-3}$ at the end of the run. In the hybrid run, by contrast, each
compression drives a shock that converts bulk kinetic energy into heat; the
resulting thermal pressure both damps successive bounces and drains the
pulsation energy, so the oscillations damp within $\sim 5$--$8\,$ms and the star
settles smoothly onto its new equilibrium. Another channel to further increase the damping of the curves is mass shedding (which only occurs in the hybrid runs), when matter is shed from the star, kinetic energy from the pulsations is lost and transferred to the outgoing matter, thus making the oscillation amplitude decrease after each bounce, see \cite{2006MNRAS.368.1609D}. Gravitational wave emission is not expected to contribute significantly to the damping of the oscillation, since the characteristic timescales of this mechanism are longer than those seen in Fig. \ref{fig:rhoc} (see Sec. \ref{subsec:GW}, where the emission peak of the GWs occurs a few ms after the damping has already occurred). The contrast between the two curves
makes explicit the role played by shock heating in regulating the migration, and
reproduces the qualitative behavior reported for the canonical spherical
test~\cite{Font:2001_grqc_0110047}.

The absence of a thermal damping channel in the cold run is worth making
explicit, because it provides a controlled measurement of the scheme's numerical
dissipation. The inset of Fig.~\ref{fig:rhoc} shows $\rho_c(t)$ for a
\emph{non-rotating} cold model. A purely radial pulsation of a spherical star
emits no gravitational radiation, and the barotropic EoS of the cold polytrope
$p=p(\rho)$, enforced throughout the evolution, generates no entropy; in the
absence of numerical error the oscillation would therefore persist undamped
indefinitely. The slow secular decay actually observed thus measures the
numerical dissipation of the scheme, dominated by the diffusivity of the HLLE
approximate Riemann solver acting at the stellar surface, which is a contact
discontinuity swept back and forth across the grid at each pulsation. This
baseline is useful in interpreting the rotating cold runs, but it is not the
only effect there. Rotation opens genuinely physical channels for damping the
quasi-radial pulsation that are absent in the non-rotating, spherical case: the
pulsation couples to the mass quadrupole and radiates gravitational waves
(Sec.~\ref{subsec:GW}); it drives the small oscillating differential-rotation
component visible in Fig.~\ref{fig:H3_omega} below, allowing pulsation energy to leak into
rotational shear; moreover, being large-amplitude and harmonic, as evidenced by the
$2F$ and $3F$ harmonics of Fig.~\ref{fig:qrmodes}, it can transfer energy to
higher-frequency modes that damp more rapidly. The damping seen in the rotating
cold runs should therefore be read as a combination of the numerical baseline
established here and these rotation-enabled physical channels, rather than as a
purely numerical effect.

\subsubsection{Thermal offset of the migration endpoint}
\label{subsubsec:thermal_offset}

\begin{figure}[tbp]
    \centering
    \includegraphics[width=\linewidth]{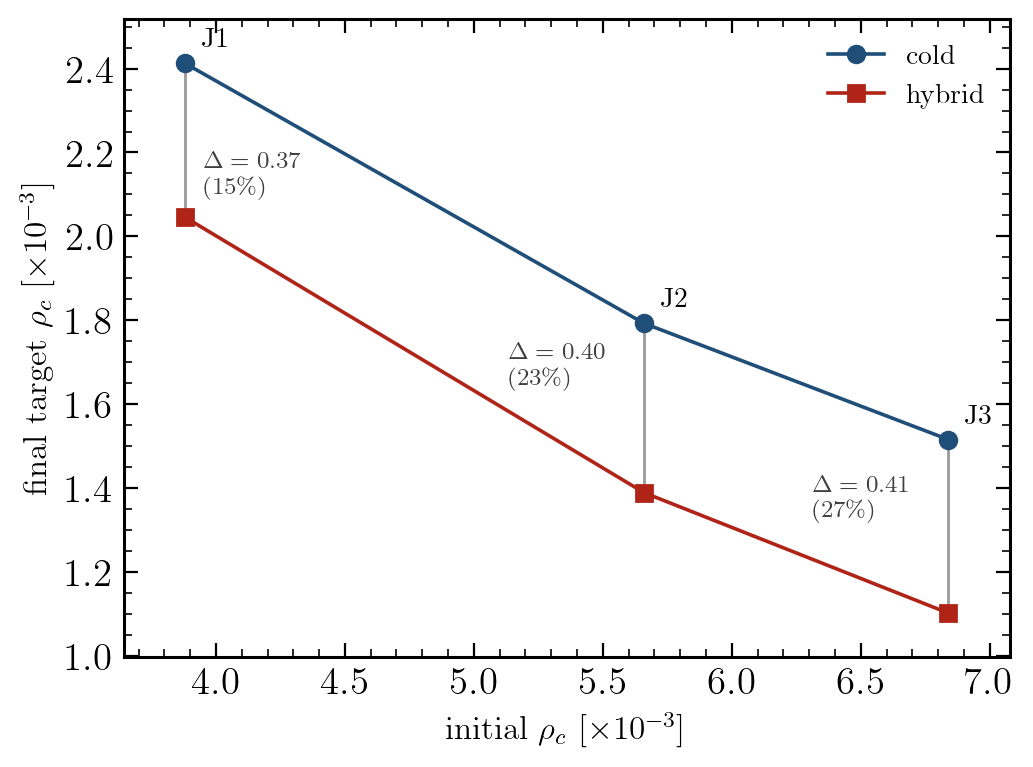}
    \caption{Final (target) central density of the migrated remnant for the
    constant-$\bar J = 0.01$ models J1, J2 and J3, evolved with the cold
    (circles) and hybrid (squares) equations of state, as a function of the
    initial central density.}
    \label{fig:rhodiff}
\end{figure}

To isolate the thermal imprint on the endpoint of the migration we compare, for
each model of the constant-$\bar J$ sequence (J1, J2, J3), the final central
density reached under the two equations of state (Fig.~\ref{fig:rhodiff}). Because
the cold pulsations persist, we define the cold endpoint as the time average of
$\rho_c$ over the final few milliseconds of the evolution. For the hybrid runs,
which settle, we take the relaxed value. Across the whole sequence the hybrid
remnant relaxes to a systematically lower central density than its cold
counterpart, by $\Delta\rho_c \simeq (0.37,\,0.40$, \,$0.41)\times 10^{-3}$ for
J1, J2 and J3, respectively. The absolute offset is thus nearly constant along
the sequence, while its fractional size grows from $\approx 15\%$ (J1) to
$\approx 23\%$ (J2) and $\approx 27\%$ (J3) as the endpoints move to lower
density.

This offset is consistent with a genuine physical effect rather than a numerical artifact. The entropy generated at the rebound shocks in the hybrid runs supplies thermal pressure that supports the remnants at lower central densities, as expected for shock-heated matter. Mass shedding may also contribute to the offset by altering the baryonic mass and angular momentum retained by each remnant. The near-constancy of the absolute shift, together with the growth of its relative importance toward the more strongly migrating models, suggests that the combined influence of thermal support and mass shedding remains comparable across the sequence while the baseline compactness on which it acts decreases.

\subsubsection{Frequency content of the post-bounce pulsations}
\label{subsubsec:qrmodes}
The post-bounce pulsations exhibit a well-defined frequency content that
provides a quantitative comparison between the two thermodynamic
prescriptions. For each evolution, we analyze the central-density
signal over the interval \(t=2\)--\(22\,\mathrm{ms}\). We linearly
detrend the signal, apply a Hann window and compute a zero-padded fast
Fourier transform. The location of the dominant spectral peak is
refined by parabolic interpolation around the largest spectral bin.
The physical frequency resolution is determined by the analyzed
duration, \(T\simeq20\,\mathrm{ms}\), and is therefore
\(\Delta f=1/T\simeq0.05\,\mathrm{kHz}\). Zero padding provides a
finer frequency grid but does not improve this physical resolution.

\begin{figure}[tbp]
    \centering
    \includegraphics[width=\linewidth]{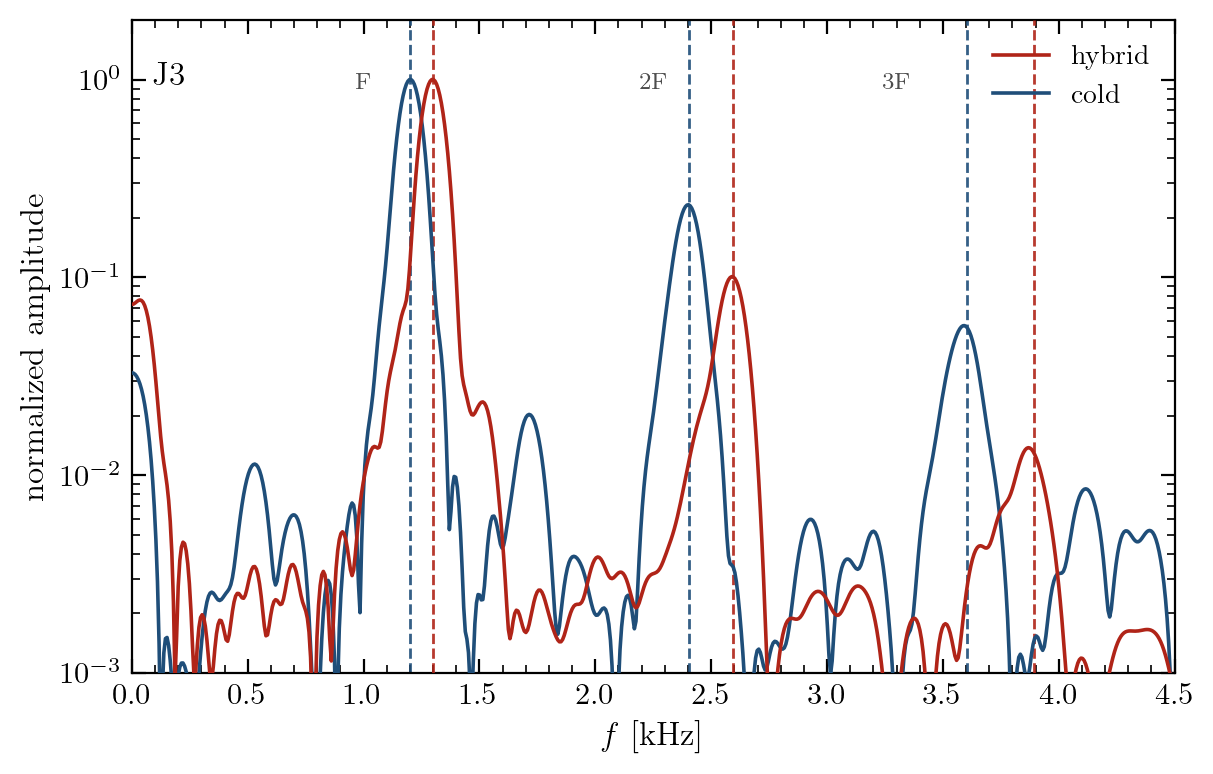}
\caption{Amplitude spectra of the post-transient central density
($\rho_c(t)$) for the J3 model evolved with the cold (blue) and hybrid
(red) equations of state, normalized separately to unit peak. Dashed
lines mark the quasi-radial fundamental (F) and its nonlinear
harmonics (2F) and (3F).}
    \label{fig:qrmodes}
\end{figure}

We interpret the dominant peak as the fundamental quasi-radial mode
\(F\) excited by the migration. Weaker peaks occur near \(2f_F\) and
\(3f_F\), with deviations from these integer multiples smaller than
the nominal frequency resolution (Fig.~\ref{fig:qrmodes}). Their
approximately harmonic spacing supports their interpretation as
nonlinear harmonics generated by the anharmonic, large-amplitude
quasi-radial oscillation, rather than as independent quasi-radial
overtones, whose eigenfrequencies are not generally integer multiples
of \(f_F\).

For J3, the estimated quasi-radial fundamental frequencies are
\(f_F\approx1.17\,\mathrm{kHz}\) in the cold evolution and
\(f_F\approx1.26\,\mathrm{kHz}\) in the hybrid evolution. The
corresponding \(2F\) and \(3F\) peaks shift consistently with the
fundamental, supporting their interpretation as nonlinear harmonics of
the same oscillation. The spectra therefore suggest that, for identical
initial data, retaining the shock-generated thermal component produces
a post-migration remnant with a higher quasi-radial frequency.

This ordering should not be regarded as a universal relation between
thermal content and pulsation frequency. The quasi-radial frequency
depends on the complete pressure, enthalpy, density and rotation
profiles of the remnant, rather than on its central density alone. In
the present hybrid prescription, the thermal component modifies the
local adiabatic pressure response and can strengthen the restoring
force in the shock-heated regions. The observed upward shift is
therefore consistent with a change in the effective stiffness of the
remnant, although differences in radius, retained mass and angular
momentum, and nonlinear oscillation amplitude may also contribute.

\subsection{Slowly rotating migration: model H1 ($r_p/r_e=0.95$)}
\label{subsec:H1}

\begin{figure*}[t]
    \centering
    \includegraphics[width=1\linewidth]{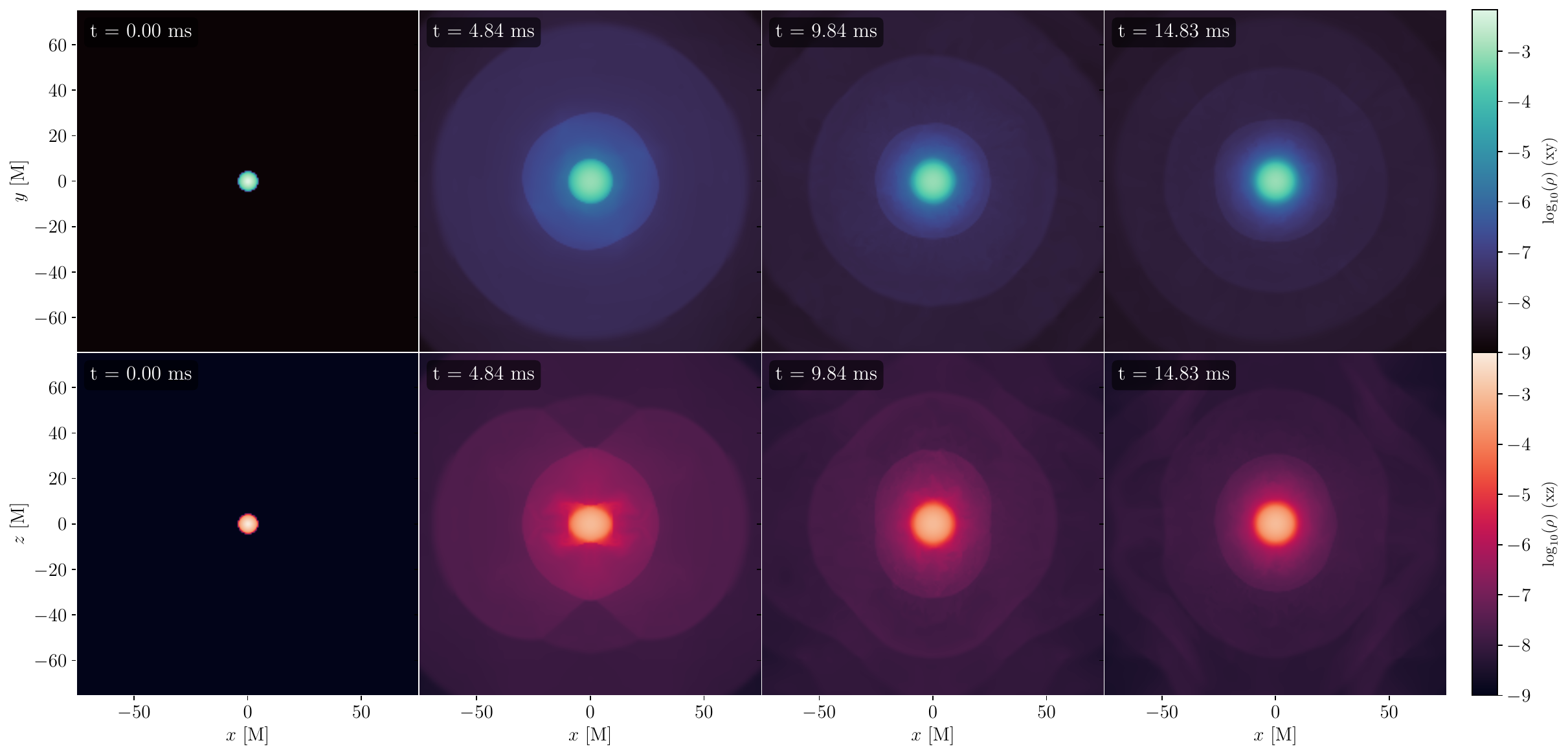}
    \caption{Model H1 ($\rho_c = 8\times 10^{-3}$, $r_p/r_e = 0.95$): evolution of
    the rest-mass density $\log_{10}\rho$ on the equatorial ($x$--$y$, top row)
    and meridional ($x$--$z$, bottom row) planes at four representative times
    (increasing left to right).}
    \label{fig:H1_rho}
\end{figure*}

Model H1 is the closest to a spherical TOV configuration and
constitutes our reference benchmark. Figure~\ref{fig:H1_rho} shows the
rest-mass density on the equatorial ($x$--$y$, top row) and meridional
($x$--$z$, bottom row) planes at four representative
times. The qualitative picture mirrors the canonical non-rotating
migration~\cite{Font:2001_grqc_0110047}: the star expands
and ejects an outgoing shock wave that sweeps the artificial atmosphere
into a rarefied halo extending well beyond the original surface. The oblateness of the initial configuration is preserved throughout
the subsequent oscillations, with small visible departure from
axisymmetry, like in the second figure from the top row of Fig. \ref{fig:H1_rho}.

\begin{figure}[tbp]
    \centering
    \includegraphics[width=\linewidth]{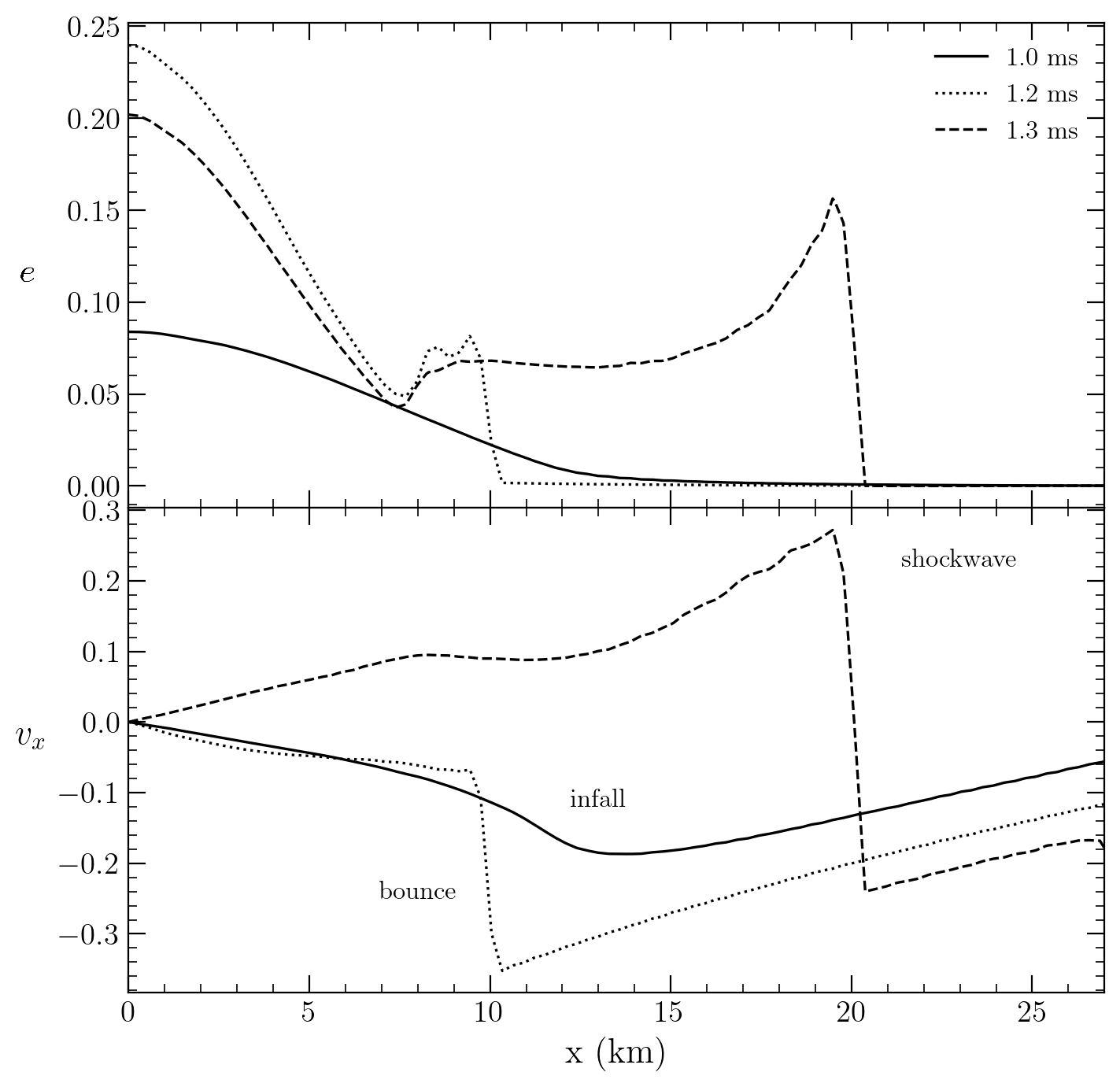}
    \caption{Model H1: radial profiles of the specific internal
    energy $e$ (top) and the $x$-velocity $v_x$ (bottom) at
    three times bracketing the first rebound. The single outgoing
    shock characteristic of the spherical migration~\cite{Font:2001_grqc_0110047} is clearly
    identifiable.}
    \label{fig:H1_wave}
\end{figure}

Figure~\ref{fig:H1_wave} resolves the radial structure of the bounce.
The top panel shows the specific internal energy $e$ and the
bottom panel the $x$-component of the three-velocity $v_x$, at three
snapshots bracketing the first rebound: $t=1.0\,$ms (still in the
\emph{infall} phase, with $v_x<0$ throughout the star), $t=1.2\,$ms
(the \emph{bounce}, marked by a near-discontinuous spike of $v_x$ at
the contact surface), and $t=1.3\,$ms (the outgoing \emph{shockwave},
already at $x\sim 20\,$km with a characteristic post-shock energy
plateau). The simple one-shock structure observed here is the
hallmark of the nearly spherical case.

\subsection{Rapidly rotating migration: model H2 ($r_p/r_e=0.80$)}
\label{subsec:H2}

\begin{figure*}[t]
    \centering
    \includegraphics[width=1\linewidth]{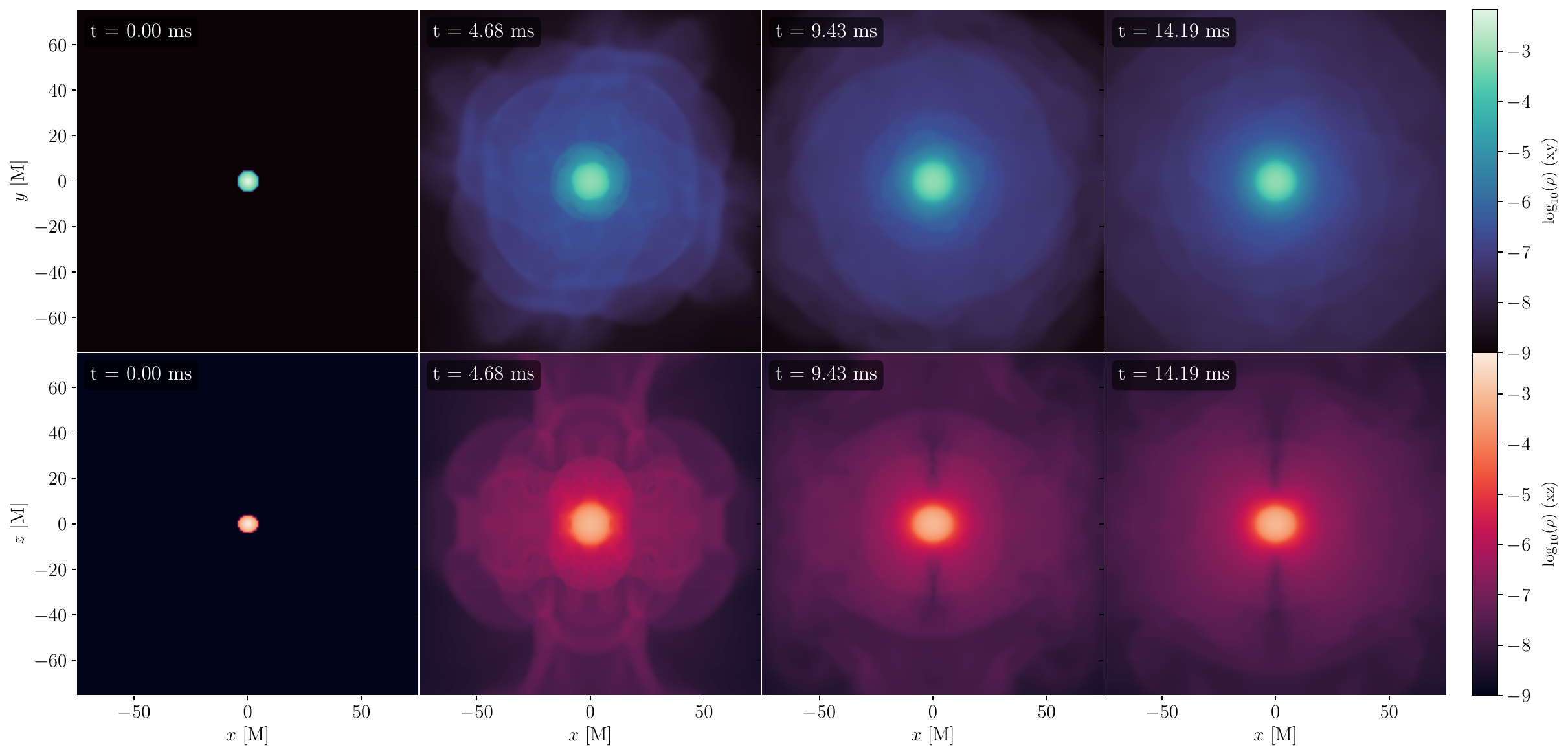}
    \caption{Same as Fig.~\ref{fig:H1_rho} for the more rapidly rotating model
    H2 ($\rho_c = 8\times 10^{-3}$, $r_p/r_e = 0.80$).}
    \label{fig:H2_rho}
\end{figure*}

Model H2 shares the same central density as H1 but rotates
substantially faster. The migration is qualitatively more violent.
The equatorial snapshots in the top row of Fig.~\ref{fig:H2_rho} display a
shock pattern in the first $\sim 5\,$ms
that subsequently relaxes to a smoother, slightly oblate
configuration. The meridional view in the bottom row of Fig.~\ref{fig:H2_rho} shows
the expected pole/equator asymmetry, where matter is preferentially expelled
along the equatorial plane.

\begin{figure}[tbp]
    \centering
    \includegraphics[width=\linewidth]{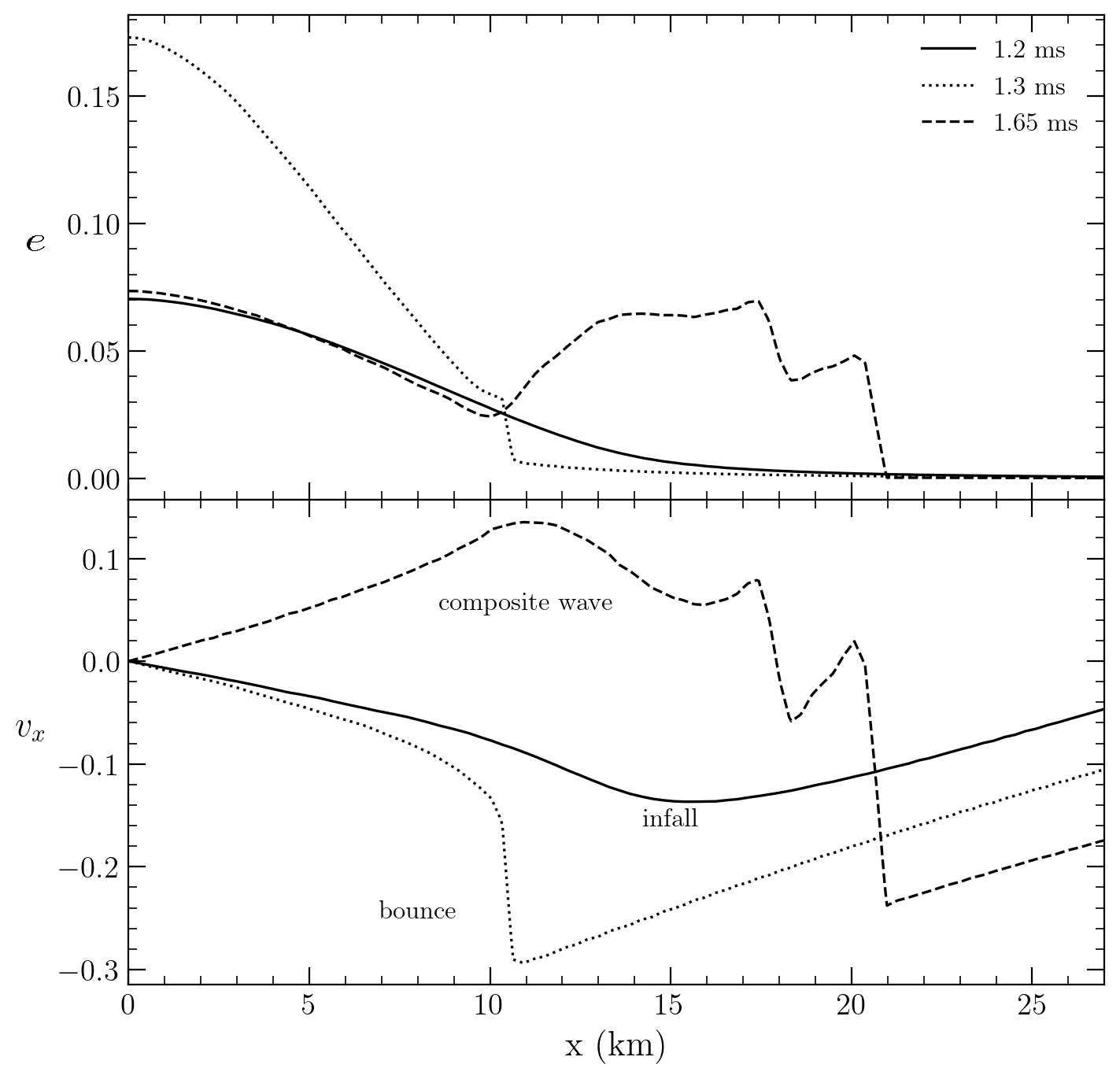}
    \caption{Model H2: radial profiles of $e$ (top) and $v_x$
    (bottom). The single shock of the slowly rotating case is replaced
    by a composite wave structure.}
    \label{fig:H2_wave}
\end{figure}

The wave analysis in Fig.~\ref{fig:H2_wave} shows a
qualitative difference with respect to H1: the single rebound shock is
replaced by a composite wave, in which a forward-propagating
shock at $x\sim 17$--$20\,$km coexists with an inward-propagating
rarefaction. This dual structure is characteristic of rotating bounces, where
the centrifugal barrier weakens the equatorial compression and allows
material at intermediate radii to overshoot in both directions before
recombining.

\subsection{Strongly nonlinear case: model H3 ($r_p/r_e=0.70$)}
\label{subsec:H3}

\begin{figure*}[t]
    \centering
    \includegraphics[width=1\linewidth]{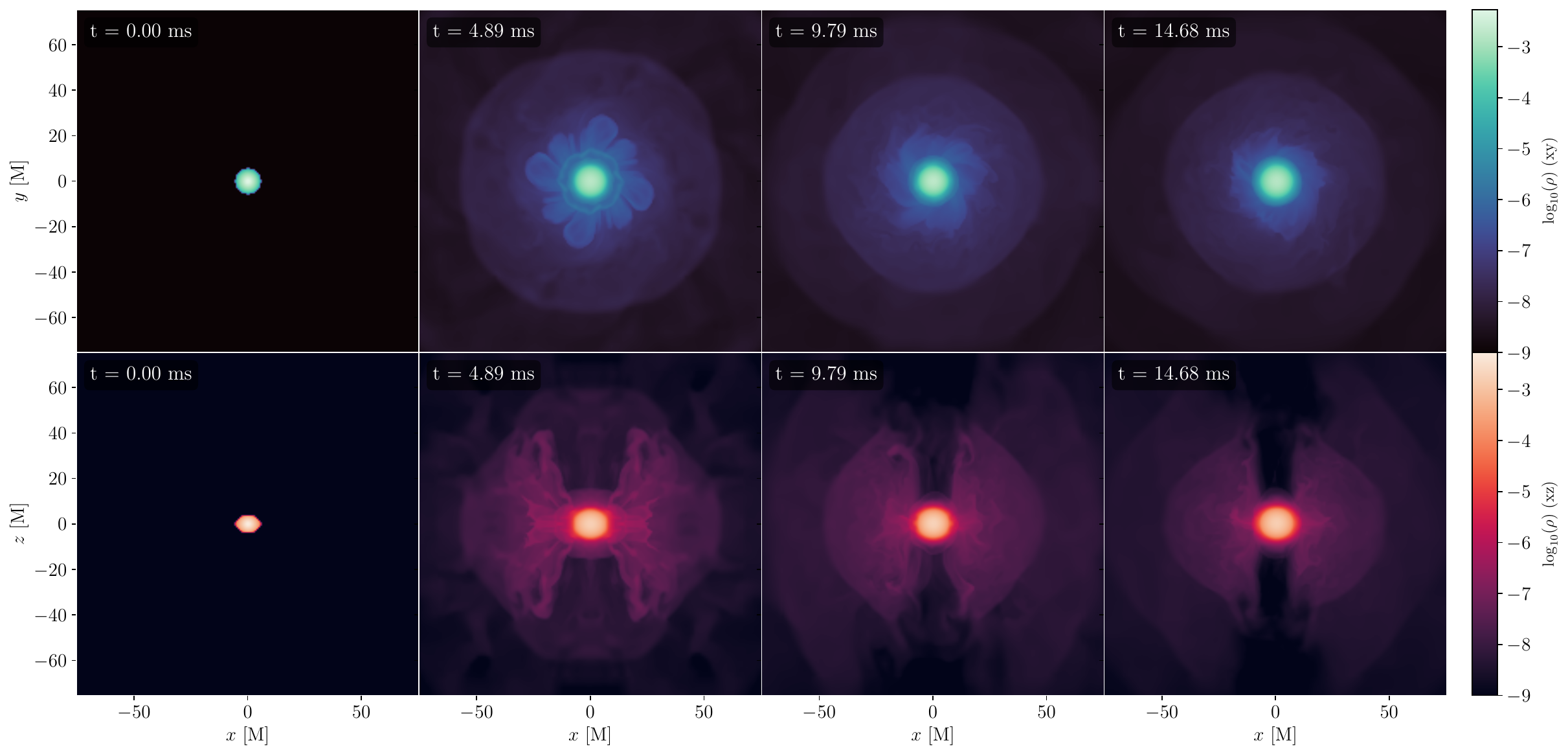}
    \caption{Same as Fig.~\ref{fig:H1_rho} for the most rapidly rotating and
    most strongly deformed model H3 ($\rho_c = 6\times 10^{-3}$,
    $r_p/r_e = 0.70$).}
    \label{fig:H3_rho}
\end{figure*}

Model H3 ($\rho_c = 6\times 10^{-3}$, $r_p/r_e = 0.70$, $\bar{J}=0.014$)
is the most rapidly rotating and most strongly deformed configuration
in our hybrid set, and it produces the most nonlinear migration. The equatorial snapshots in the top row of Fig.~\ref{fig:H3_rho} show that the first
rebound drives large-amplitude, non-axisymmetric transients in the
expanding core before the central object relaxes to a smoother,
strongly flattened state. The non-axisymmetric features are more
pronounced and longer-lived than in H2, but they remain transient: by
late times the equatorial structure has largely axisymmetrized.

The meridional view in the bottom row of Fig.~\ref{fig:H3_rho} shows the final shape
of this model. The migrated configuration is markedly oblate: matter
accumulates in an extended equatorial belt while the polar regions are
emptied, so that the post-bounce object
resembles a flattened, disk-like envelope around the dense core rather
than the mildly oblate spheroid of H1. The pole-to-equator density
contrast is substantially larger than in H2, reflecting the stronger
centrifugal barrier at this rotation rate, which suppresses equatorial
compression and channels the ejected material preferentially away from the rotation axis.

\begin{figure}[tbp]
    \centering
    \includegraphics[width=\linewidth]{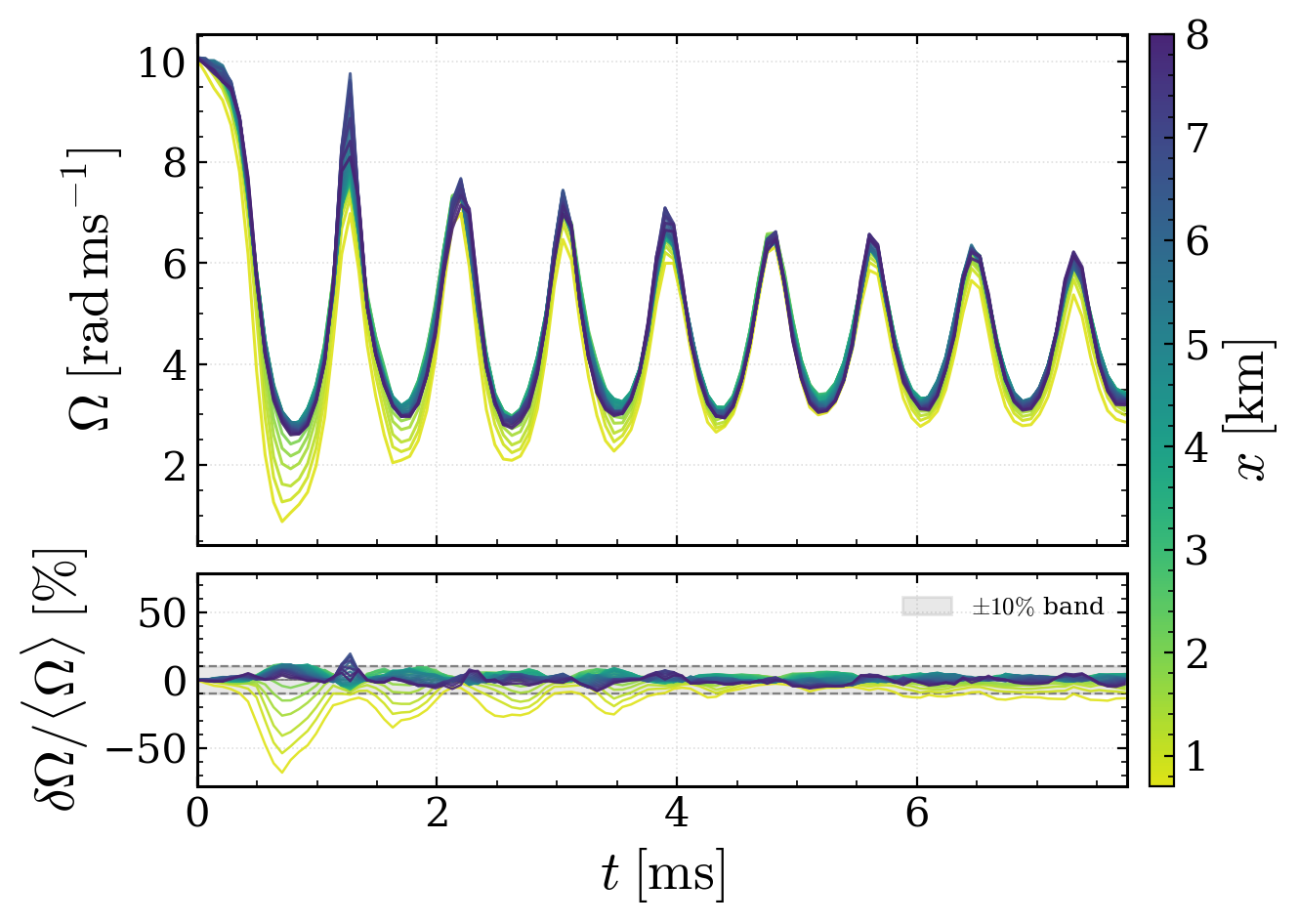}
    \caption{Equatorial angular velocity of model H3 as a
    function of time. \emph{Top:} $\Omega$ evaluated at a set of fixed
    cylindrical radii spanning the bulk of the star (colour bar), obtained
    from the coordinate angular velocity
    $\Omega = (\alpha V^{y}-\beta^{y})/x$ on the equatorial plane. \emph{Bottom:} the deviation of each curve from the mean
    over the sampled radii, $\delta\Omega/\langle\Omega\rangle$.}
    \label{fig:H3_omega}
\end{figure}

The long-term behavior of the rotation law is summarized in
Fig.~\ref{fig:H3_omega}. The initial \texttt{RNS} model is a uniformly
rotating equilibrium, so the profile is flat across the star at $t=0$ and
drops sharply at the stellar surface. As the star migrates, it expands by
roughly a factor two in equatorial radius, and the angular velocity
of the bulk falls accordingly, from $\Omega\simeq10$ to
$\simeq3\,$rad\,ms$^{-1}$, where $\Omega$ is calculated from
\begin{equation}
    \Omega=\frac{u^\phi}{u^t}.
\end{equation}
This is consistent with the larger moment of
inertia of the migrated configuration at fixed angular momentum.
Superimposed on this secular spin-down, $\Omega$ oscillates with
large amplitude, tracking the quasi-radial pulsation of the remnant: the
star spins up as it contracts and down as it expands, so that the rotation
rate is linked to the breathing motion rather than evolving independently.
 
\begin{figure}[tbp]
    \centering
    \includegraphics[width=\linewidth]{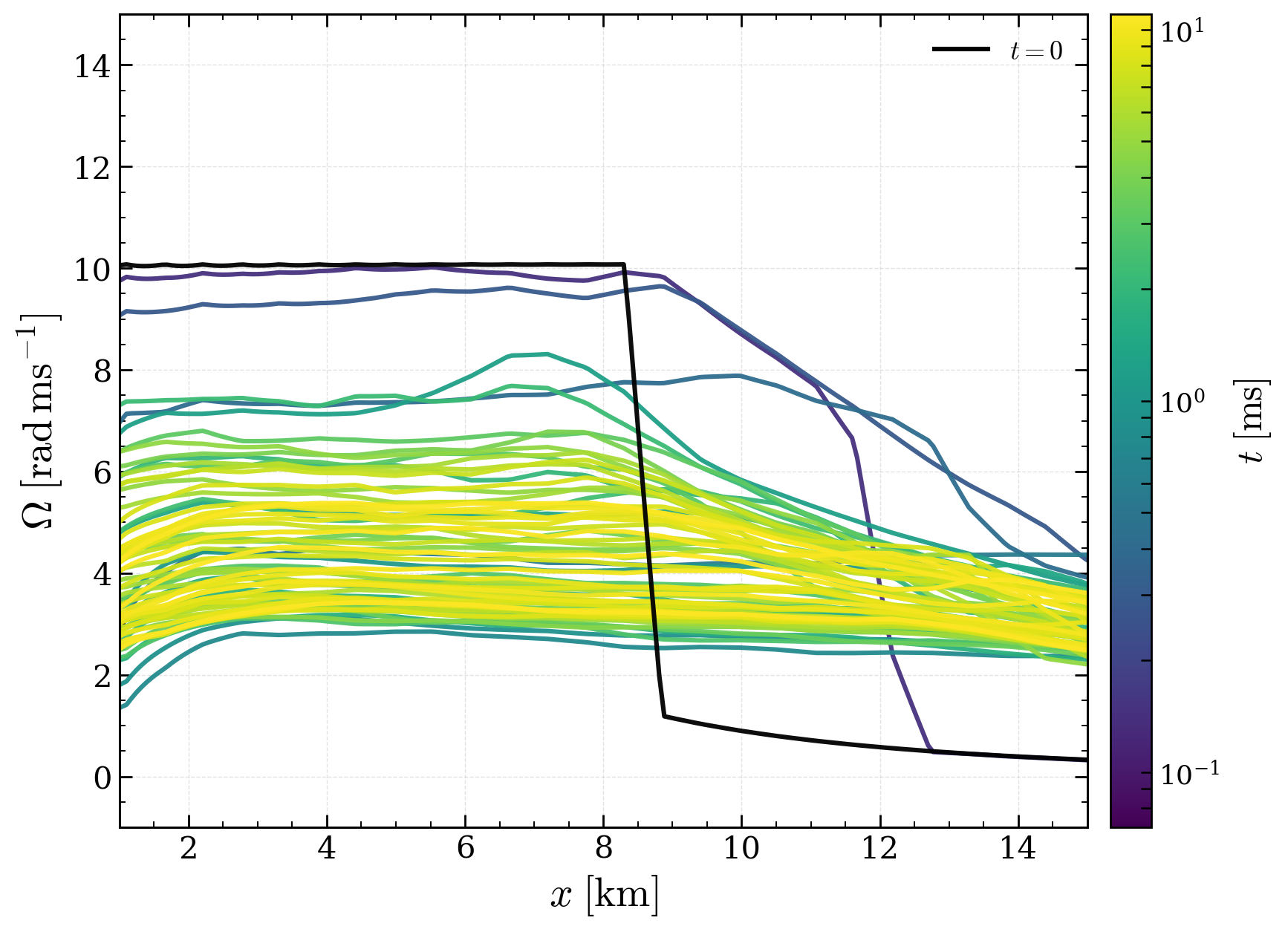}
    \caption{Equatorial angular-velocity profiles of model H3 at selected times during the migration. The initial uniformly rotating configuration is shown in black.
}
    \label{fig:H3_omega_profiles}
\end{figure}

Although the
rebound and the associated shock heating are violent, the remnant does not
settle into a strongly differentially rotating state: after the transition the
angular velocity is the same, to within $10\%$, across most of the bulk of the star (lower panel of Fig.~\ref{fig:H3_omega}). The residual
differential rotation is itself modulated at the pulsation frequency,
peaking near maximum expansion, when the outer layers lag the core because
their fractional change in radius is the larger; it relaxes again on
compression. The migrated configuration is therefore compatible with a stable uniformly rotating equilibrium state after the transition (see Fig.~\ref{fig:H3_omega_profiles}). This is a nontrivial outcome for
a benchmark: it means that persistent strong differential rotation could indicate excessive numerical angular-momentum transport rather than
capturing a physical channel.
 
We note in passing that extracting $\Omega$ near the rotation axis requires
care. We use the projection along the $x$-axis $\Omega=(\alpha V^{y}-\beta^{y})/x$, this presents a $0/0$ limit as
$x\to 0$, so any component of the numerator that does not vanish linearly
through the origin, a small drift of the configuration off the
coordinate centre, or a nonzero shift at the origin developed by the
$\Gamma$-driver appears as a spurious $c/x$ contribution and mimics a
rapidly spinning core. Symmetrizing $\Omega$ about the axis removes the
odd-$x$ part of this contamination.

\subsection{Gravitational-Wave Emission}
\label{subsec:GW}
 
\begin{figure}[tbp]
    \centering
    \includegraphics[width=\linewidth]{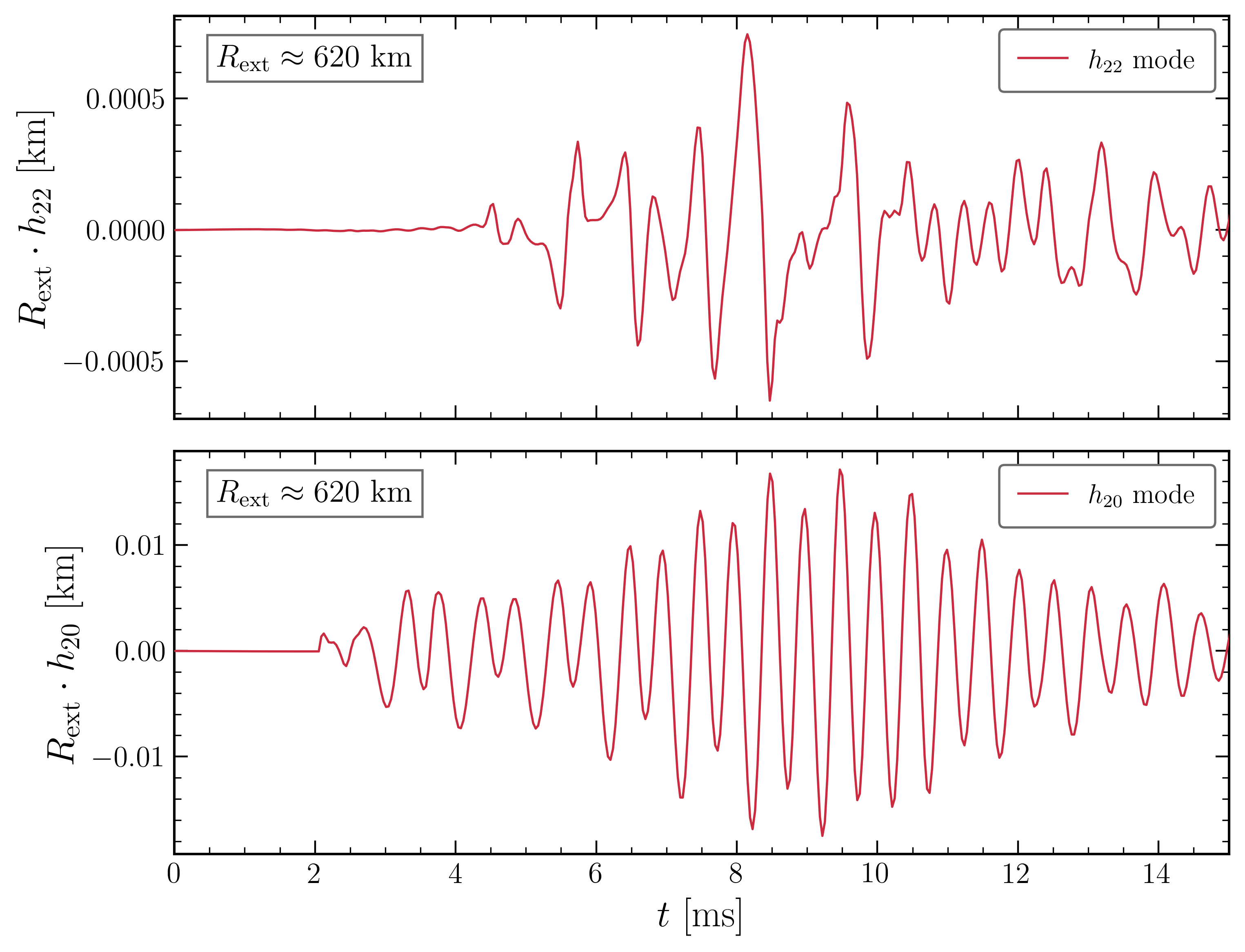}
    \caption{Time evolution of the gravitational-wave strain scaled by radius for
model H1, extracted at $R_{\rm ext}\approx620\,$km: the $h_{22}$ mode (top) and
the $h_{20}$ mode (bottom).}
    \label{fig:gwextraction2220}
\end{figure}

We now examine the gravitational-wave signal accompanying the migration,
using model H1 as the reference case. We first describe the time-domain
strain in the two dominant multipoles, $h_{22}$ and $h_{20}$
(Fig.~\ref{fig:gwextraction2220}), and then analize their frequency
content and its relation to the quasi-radial pulsation identified in
Sec.~\ref{rhocsection} (Fig.~\ref{fig:fft20}). Both multipoles are
extracted at $R_{\rm ext}\approx 620\,$km, for this purpose we use larger domains in these simulations while keeping the resolution of the neutron star the same. Because the comparison with the
central density is made within a single evolution, the $\rho_c$ spectrum
shown in this section is that of the same H1 run rather than the
constant-$\bar J$ models of Sec.~\ref{rhocsection}.

We decompose \(\Psi_4\) into spin-weighted spherical-harmonic modes on coordinate extraction spheres. At each extraction radius, we estimate the asymptotic waveform using the nonspinning second-order perturbative extrapolation formula of Ref.~\cite{Nakano:2015}, including corrections through second order in the inverse areal radius. This procedure is applied independently at each extraction sphere and should not be confused with a polynomial fit across several extraction radii. We obtain the strain modes \(h_{\ell m}\) by integrating the corrected \(\Psi_4^{\ell m}\) twice in the time domain, removing the resulting linear drift, and applying a zero-phase Butterworth high-pass filter. We present the asymptotic estimate obtained from the outermost extraction sphere, whose coordinate radius is \(R_{\rm ext}=420\,M_\odot\simeq620\,\mathrm{km}\), as \(r_{\rm areal}h_{\ell m}\).

The amplitude of the signal is small: the radius-scaled $h_{22}$ strain
remains of order $\sim10^{-4}\,$km throughout. This is physically
consistent with the dynamical scenario. Although the migration involves a
violent structural transition of the core, with the central density varying
by more than an order of magnitude, the event is neither very massive nor
strongly asymmetric when compared with, for instance, the merger of a
binary system, and the quadrupole moment that rotation makes available to
the radiation field is correspondingly small.
 
A characteristic feature of the waveform is the appearance of the
first signal near \(t\approx2\,\mathrm{ms}\). This early contribution
is associated with the onset of the migration and the initial expansion
of the rotating star at \(t\approx0\), rather than with the subsequent
hydrodynamic rebound. The light-travel time from the stellar interior
to the extraction sphere is approximately
\(\Delta t=R_{\rm ext}/c\simeq2.06\,\mathrm{ms}\), consistent with the
arrival time of this initial response. Following these early transients, the waveform exhibits
sustained oscillations associated with the pulsating post-migration
configuration.

The relative strength of the two channels reflects the geometry of the transition. The migration of the slowly rotating H1 model remains approximately axisymmetric, and the \(l=2,m=0\) strain exceeds the non-axisymmetric \(l=2,m=2\) component by a factor of approximately \(20\)--\(30\) (Fig.~\ref{fig:gwextraction2220}). The \(h_{20}\) envelope grows during the first several milliseconds, reaches its maximum around \(t\approx8\)--\(9\,\mathrm{ms}\), and subsequently rings down. The initial stellar motion is predominantly quasi-radial; however, rotation couples this motion to an axisymmetric quadrupolar deformation, allowing the quasi-radial frequency to appear in \(h_{20}\). The waveform also contains an axisymmetric quadrupolar \(f\)-mode of the post-migration remnant. These contributions are distinguished through the spectral comparison presented below.
 
\subsubsection{Frequency content and the quasi-radial mode}
\label{subsubsec:GWspectra}
 
\begin{figure}[tbp]
    \centering
    \includegraphics[width=\linewidth]{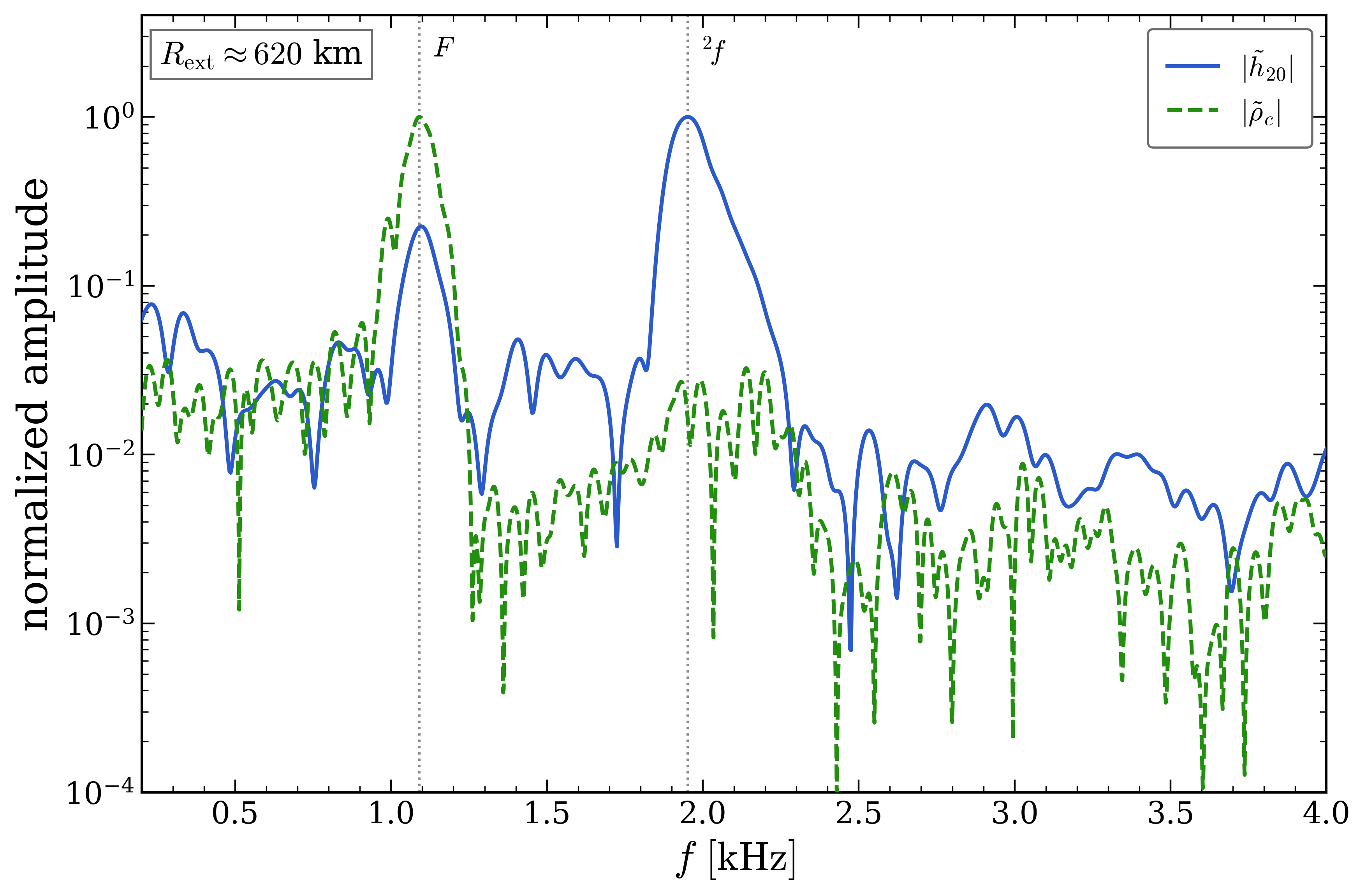}
    \caption{Spectra of the gravitational-wave strain and of the central density
for model H1, each normalized to its own peak. The figure shows the amplitude spectrum of the $l=2$, $m=0$ strain $h_{20}$ extracted at
$R_{\rm ext}\approx620\,$km (solid), overlaid with that of the central
rest-mass density $\rho_c(t)$ of the same evolution (dashed).}
    \label{fig:fft20}
\end{figure}

Figure~\ref{fig:fft20} compares the spectrum of the strain with that of the
central density. Both time series are linearly detrended and tapered with a
Hann window before a zero-padded fast Fourier transform,
following the same procedure applied to $\rho_c(t)$ in
Sec.~\ref{rhocsection}; the $\rho_c$ record is analyzed only for
$t>3\,$ms, after the initial expansion and first bounces, so that the
secular settling ramp does not dominate the low-frequency end of the
spectrum. Zero padding refines the frequency grid but not the resolution,
which is set by the analyzed duration as
$\Delta f \simeq 1/T \approx 0.03\,$kHz.
 
The $h_{20}$ spectrum resolves two distinct peaks (Fig.~\ref{fig:fft20}). The lower one, at $f_F\approx1.09\,$kHz for this H1 run (not to be confused
with the J3 values quoted in Sec.~\ref{rhocsection}, which refer to a different
model), coincides to within the spectral resolution with the fundamental
quasi-radial peak $F$ of the $\rho_c$ spectrum.
This coincidence closes the argument anticipated in Sec.~\ref{rhocsection}: a
purely radial pulsation of a spherical star emits no gravitational radiation,
but rotation breaks that degeneracy by coupling the quasi-radial mode to the
mass quadrupole, so the same oscillation that appears as a compression and
rarefaction of the core in $\rho_c(t)$ is radiated through the axisymmetric
$h_{20}$ channel. The same coupling is visible in the time domain in the
rotation profile of Sec.~\ref{subsec:H3}, where the bulk angular velocity rises
and falls in phase with the breathing motion of the star
(Fig.~\ref{fig:H3_omega}).
 
The dominant \(h_{20}\) peak lies at \(f\simeq1.93\,\mathrm{kHz}\) and appears exclusively in the gravitational-wave spectrum. This spectral selectivity is characteristic of a non-radial oscillation, since regular \(l\geq1\) density perturbations vanish at the stellar center. Its separation from the expected harmonic frequency, \(2f_F\simeq2.18\,\mathrm{kHz}\), is larger than the spectral resolution. Together with its appearance in the \(l=2,m=0\) channel, these properties support its interpretation as a candidate axisymmetric quadrupolar \(f\)-mode of the migrated remnant.

For model H1, the central-density and \(h_{20}\) spectra independently recover the same quasi-radial frequency, \(f_F\simeq1.09\,\mathrm{kHz}\). This agreement supports the identification of the lower-frequency \(h_{20}\) peak with the fundamental quasi-radial mode and provides a characteristic spectral feature for reproductions of the migration test. The oscillations observed in the rotation profiles of model H3 complement this result by illustrating the coupling between the breathing motion and the rotational dynamics. In the present analysis, we use \(f_F\) as a characteristic observable of the migrated remnant; relating it quantitatively to the neutral-stability line would require dedicated mode calculations along the corresponding rotating equilibrium sequence.

\section{Conclusions}
\label{sec:conclusions}

We have extended the standard neutron-star migration test from spherical
models~\cite{Font:2001_grqc_0110047} to uniformly rotating
configurations evolved in three spatial dimensions. The simulations
combine the \texttt{GRHayL} hydrodynamics infrastructure with the
\texttt{ML\_BSSN} spacetime solver and employ either a cold polytropic EoS or a hybrid cold$+$thermal prescription.
Evolving identical initial data with these two prescriptions allows us
to isolate the influence of shock-generated thermal pressure on the
migration dynamics and final state.

For the cold-EoS evolutions, dynamically unstable models move toward their analogue stable configurations associated with their
initial angular momentum in the
$\bar{M}_0$--$\bar{\epsilon}_c$ plane. This behavior extends the
familiar spherical migration test to rotating configurations and
provides a direct diagnostic of whether angular momentum is conserved
with sufficient accuracy during the evolution.

Rotation also changes the morphology of the rebound. The slowly
rotating model H1 approximately retains the main features of the spherical
case~\cite{Font:2001_grqc_0110047}: a single outgoing shock followed by almost
radial pulsations that are damped by shock heating. At higher rotation
rates, models H2 and H3 develop a composite
forward-shock/inward-rarefaction structure, preferential equatorial
ejection, and transient non-axisymmetric features during the early
post-bounce evolution. These effects are absent from the spherical test
and make rotating migration a more demanding test of the coupled
hydrodynamic and spacetime evolution.

Despite the violent rebound, the migrated remnant does not develop a
persistent strongly differential rotation profile. After the initial
transient, the angular velocity remains approximately uniform across
the bulk of the star, with typical deviations of $10\%$. The
remaining deviations oscillate at the quasi-radial frequency: the star
spins up during compression and spins down during expansion about a
mean angular velocity that decreases as the remnant expands. This
behavior indicates that the migration primarily changes the moment of
inertia of the star without producing a long-lived differentially
rotating state.

The comparison between the cold and hybrid evolutions reveals a
systematic thermal shift of the migration endpoint. Along the
constant-$\bar{J}$ sequence, the shock-heated remnants settle at lower
central densities than their cold counterparts, with
$\Delta\rho_c\simeq0.4\times10^{-3}$ for the models considered here.
Although the absolute displacement is approximately constant, its
fractional magnitude increases from about $15\%$ to $27\%$ toward the
more strongly migrating models. The frequency-domain analysis shows hybrid runs oscillate at higher frequencies. Neglecting thermal
pressure would therefore place the endpoint at a higher central density
and predict a lower quasi-radial frequency.

The gravitational-wave signal provides an independent diagnostic of
the post-bounce dynamics. For model H1, the axisymmetric $h_{20}$
component dominates the non-axisymmetric $h_{22}$ component. The
$h_{20}$ spectrum contains a lower-frequency peak coincident, within
the spectral resolution, with the quasi-radial fundamental identified
in the central-density evolution. It also contains a stronger
higher-frequency peak consistent with the quadrupolar $f$-mode of the
migrated remnant. The arrival of the initial signal is consistent with the
propagation time from the stellar core to the extraction sphere,
providing a basic check of the waveform extraction.

Taken together, these results motivate a rotating extension of the
standard migration benchmark. The benchmark does not require an unstable model to migrate rather than collapse, since the selected branch depends on the direction of the applied or numerically induced perturbation. Instead, it specifies the expected behavior conditional on migration occurring. For the initial models and numerical
setup specified in this work, a successful reproduction should exhibit
the following qualitative behavior: (i) the projected cold-EoS migration final states
should lie on the constant-$\bar{J}$ equilibrium sequences
associated with the initial models
(Figs.~\ref{fig:Meps_constrho}--\ref{fig:Meps_constJ}); (ii) the central
density should undergo a deep first rebound followed by weakly damped
pulsations for the cold EoS and shock-mediated damping for the hybrid
EoS (Fig.~\ref{fig:rhoc}); (iii) at fixed initial baryonic mass and
angular momentum, thermal pressure should shift the migration endpoint
toward lower central density (Fig.~\ref{fig:rhodiff}); and (iv) the
rebound should change from a predominantly single outgoing shock at
low rotation to a composite wave structure at higher rotation,
accompanied by preferential equatorial ejection and transient
non-axisymmetric structure. Reproducing these features would provide a
joint test of the conservative-to-primitive recovery, HRSC
reconstruction, angular-momentum conservation, and spacetime evolution
during large density excursions and strong shocks.

Natural extensions include models closer to the mass-shedding limit,
for which persistent non-axisymmetric deformations may become
important, and the inclusion of magnetic fields, finite-temperature or
tabulated equations of state, and neutrino transport. These additions
would bring the benchmark closer to the conditions encountered in
post-merger remnants, although they would reduce the clean separation
between hydrodynamic and thermal effects afforded by the cold/hybrid
comparison employed here.

The numerical data supporting the findings of this study are available
from the corresponding author upon request.

\section*{Acknowledgments}
OHP acknowledges support by the Generalitat Valenciana and the European Social Fund (ESF) CIACIF 475/2024. This research is supported by the Spanish Agencia Estatal de Investigación (grant PID2024-159689NB-C21) funded by MICIU/AEI/10.13039/501100011033 and by FEDER / EU, by the Generalitat Valenciana (Prometeo Excellence Programme grant CIPROM/2022/49), and by the European Horizon Europe staff exchange (SE) programme HORIZON-MSCA-2021-SE-01 (Grant No.-NewFunFiCO-101086251). NS acknowledges support by the project “3rd Call for H.F.R.I. Research Projects to support Faculty Members and Researchers” (Project No.~26254) and by the European Union’s Horizon Europe Research and Innovation Programme under Grant Agreement No 101131928. The simulations were performed on the MareNostrum supercomputer at the Barcelona Supercomputing Center (BSC), through allocations granted by the Red Española de Supercomputación (RES, AECT-2026-1-0040).

\bibliographystyle{apsrev4-2}
\bibliography{bib}

\end{document}